\documentclass[]{aa}  

\usepackage{graphicx}
\usepackage{txfonts}
\usepackage{indentfirst}
\usepackage{natbib}
\usepackage{textcomp}
\usepackage{ulem}
\bibpunct{(}{)}{;}{a}{}{,} 
\usepackage[scriptsize]{subfigure}
\usepackage{amsmath1}

\usepackage{color}

\begin{document}
  
  \title{Radiation hydrodynamics with adaptive mesh refinement and application to prestellar core collapse.}
    \subtitle{I. Methods}
  
   \author{B. Commer\c con
           \inst{1,2,3,4}
           ,
            R. Teyssier\inst{2,5}
            ,
	  E. Audit\inst{2}
	  ,
           P. Hennebelle\inst{3}
 	  ,
 	  \and
	  G. Chabrier\inst{4,6}
          }
   \offprints{B. Commer\c con}

   \institute{Max Planck Institut f\"ur Astronomie, K\"onigstuhl 17, 69117 Heidelberg, Germany\\
              \email{benoit@mpia-hd.mpg.de}
         \and 
	 Laboratoire AIM, CEA/DSM - CNRS - Universit\'e Paris Diderot,
IRFU/SAp, 91191 Gif sur Yvette, France 
         \and
             Laboratoire de radioastronomie (UMR 8112 CNRS), \'Ecole Normale Sup\'erieure et Observatoire 
de Paris, 24 rue Lhomond, 75231 Paris Cedex 05, France
\and
\'Ecole Normale Sup\'erieure de Lyon, Centre de recherche Astrophysique de Lyon (UMR 5574 CNRS), 
46 all\'ee d'Italie, 69364 Lyon Cedex 07, France
\and
Universit\"at Z\"urich, Institut f\"ur Theoretische Physik,
Winterthurerstrasse 190, CH-8057 Z\"urich, Switzerland
\and
School of Physics, University of Exeter, Exeter, UK EX4 4QL\\
             }

   \date{Received October 7th, 2010; accepted February 14th, 2011}

  \abstract  {Radiative transfer has a strong impact on the collapse and the fragmentation of prestellar dense cores. }   
  {We present the radiation-hydrodynamics (RHD) solver we designed for the \ttfamily{RAMSES} \rm code. The method is designed for astrophysical purposes, and in particular for protostellar collapse.}
  { We present the solver, using the co-moving frame to evaluate the radiative quantities.  We use the popular  flux-limited diffusion approximation under the grey approximation (one group of photons). The solver is based on the second-order Godunov scheme of  \ttfamily{RAMSES} \rm for its hyperbolic part and on an implicit scheme for the radiation diffusion and the coupling between radiation and matter. }
      {We report in detail our methodology to integrate the RHD solver into  \ttfamily{RAMSES}\rm. We successfully  test the method in several conventional tests. For validation in 3D, we perform calculations of the collapse of an isolated 1 M$_\odot$ prestellar dense core without  rotation. We successfully compare the results with previous studies that used different models for radiation and hydrodynamics.}  
       {We have developed a full radiation-hydrodynamics solver in the  \ttfamily{RAMSES} \rm code that handles adaptive mesh refinement grids. The method is a combination of an explicit scheme and an implicit scheme accurate to the second-order in space. Our method is well suited for star-formation purposes. Results of multidimensional dense-core-collapse calculations with rotation are presented in a companion paper.}   

\keywords {hydrodynamics, radiative transfer - Methods: numerical- Stars:  low mass, formation - ISM: kinematics and dynamics, clouds}

\titlerunning{}
\authorrunning{B. Commer\c con et al.}
   \maketitle


\section{Introduction}

Within recent years, star-formation calculations have undergone a rapid increase in the variety of the physical models that are included. The coupling between radiative transfer and hydrodynamics has been widely studied for many year and considering different regimes and frames \citep[e.g.][]{Mihalas_book,Lowrie_2001,Mihalas_Auer_2001,Krumholz_et_al_2007}. 
 Radiation hydrodynamics (RHD) methods have been developed in grid-based codes \citep{Stone_Norman_92,Hayes_Norman_03,Krumholz_07,Kuiper_2010,Sekora_2010, Tomida_2010} and also in smoothed particles hydrodynamics (SPH) codes \citep{Boss_et_al_2000, Whitehouse_Bate_2006,Stamatellos_et_al_2007A&A}. Most of these studies use the popular flux-limited diffusion approximation \citep[FLD, e.g.][]{Minerbo_1978JQSRT,Levermore_Pomraning_1981ApJ} approximation to model the radiation transport.

 In star-formation calculations, the easiest method to take into account radiative transfer is to use a barotropic approximation, which reproduces approximately the thermal behaviour of the gas during the collapse. However, more accurate RHD calculations show  that a barotropic equation of state (EOS) cannot account for realistic cooling and heating of the gas \citep[e.g.][Commer\c con et al. in prep, hereafter Paper II]{Boss_et_al_2000,Attwood_2009}. Recently, using radiation-magnetohydrodynamics calculations, \cite{Commercon_2010L} have shown that the barotropic approximation cannot properly account for the combined effects of magnetic field and radiative transfer in the first collapse and in the first core formation. On larger scales, radiative transfer has been found to greatly reduce the fragmentation because of the radiative feedback owing to accretion and protostellar evolution \citep{Bate_2009,Offner_2009}. 

In this study, we present a new RHD solver based on the FLD approximation, which we integrate in the adaptive mesh refinement (AMR) code {\ttfamily RAMSES} \citep{teyssier-2002}. The solver is consistently integrated in the second-order predictor-corrector Godunov scheme of {\ttfamily RAMSES}, which we modify to account for the radiative pressure. But we add an implicit solver to handle the radiation diffusion and the coupling between matter and radiation, which involves physical processes on timescales much shorter than the hydrodynamical one.
The FLD is easier to implement in an AMR code than more sophisticated methods that would require an additional equation on the first moment of the radiative transfer equation \citep[e.g. M1 model, ][]{Gonzalez_2007}.
The extension to (ideal) MHD flows presents no particular difficulties and has already been used \citep{Commercon_2010L} based on the solver presented in \cite{Fromang_2006} and \cite{Teyssier_2006}.

The paper is organized as follows:  in Sect. 2 we recall the RHD equations in the comoving frame we use and we  briefly present the FLD approximation. The RHD solver for the {\ttfamily RAMSES} code is presented in Sect. 3. In Sect. 4 the method is tested for well-known test cases. As a final test, RHD dense core collapse calculations with a very high resolution are performed. Section 5 summarizes our work and the main results andr present our assessment of the method's potential.

\section{Radiation hydrodynamics in the flux-limited diffusion approximation}

\subsection{Radiation hydrodynamics in the {\it comoving} frame}

We consider the equations governing the evolution of an inviscid, radiating fluid, where radiative quantities are estimated in the comoving frame and are frequency-integrated \citep{Mihalas_book}
{\small
\begin{equation}
 \left\{
\begin{array}{ccccl}
\partial_t \rho & + & \nabla \left[\rho\textbf{u} \right] & = & 0 \\
\partial_t \rho \textbf{u} & + & \nabla \left[\rho \textbf{u}\otimes \textbf{u} + P \mathbb{I} \right]& =&  \sigma_
\mathrm{R}\textbf{F}_\mathrm{r}/c \\
\partial_t E & + & \nabla \left[\textbf{u}\left( E + P \right)\right] &= & \sigma_\mathrm{R}\textbf{F}_\mathrm{r}/c\cdot\textbf{u}
- \sigma_\mathrm{P}(4\pi B -cE_\mathrm{r})\\
\partial_t E_\mathrm{r} & + & \nabla \left[\textbf{u}E_\mathrm{r}\right] & = & -\nabla\cdot\textbf{F}_\mathrm{r} -  \mathbb{P}_\mathrm{r}:\nabla\textbf{u} + \sigma_\mathrm{P}(4\pi B -c E_\mathrm{r})\\
\partial_t \textbf{F}_\mathrm{r} & + & \nabla \left[\textbf{u}\textbf{F}_\mathrm{r}\right] & = & -c^2\nabla\cdot\mathbb{P}_\mathrm{r} - 
\left(\textbf{F}_\mathrm{r}\cdot\nabla\right)\textbf{u}  -\sigma_\mathrm{F} c\textbf{F}_\mathrm{r},
\end{array}
\right.
\label{syst1}
\end{equation}}
\noindent where $\rho$ is the material density, $\textbf{u}$ is the velocity, $P$ the thermal pressure, $\sigma_
\mathrm{R}$ is the Rosseland mean opacity,  $\textbf{F}_\mathrm{r}$ is the radiative flux, $E$ the fluid total energy $E=\rho\epsilon +1/2\rho u^2$ ($\epsilon$ is the internal specific energy), $\sigma_\mathrm{P}$ is the Planck opacity, $B=B(T)$ is the Planck function, $E_\mathrm{r}$ is the radiative energy  and $\mathbb{P}_\mathrm{r}$ is the radiation pressure. 
We see that the term $\sigma_
\mathrm{R}\textbf{F}_\mathrm{r}/c$ acts as a radiative force on the material. The material energy lost by emission is transferred into radiation, and radiative energy lost by material absorption is added to the material. To close this system, we need two closure relations: one for the gas and one for the radiation. In this work, we only consider an ideal gas closure relation for the material: $P=(\gamma -1)e=\rho k_\mathrm{B}T/\mu m_\mathrm{H}$ where $\gamma$ is the specific heats ratio, $\mu$ is the mean molecular weight, and $e=\rho c_v T$ is the gas internal energy. For the radiation, we use the FLD approximation to close the system of moment equations \citep{Minerbo_1978JQSRT,Levermore_Pomraning_1981ApJ}. In this work, we consider only the simplified case of a {\it grey material}, where all frequency-dependent quantities are integrated over frequency. We cannot use a frequency-dependent model for our purpose because of CPU limitation. 

In comparison with the laboratory frame formulation, \cite{Castor_1972} demonstrated that in the comoving frame, an additional advective flux of the radiation enthalpy is not taken into account. In the dynamic diffusion regime, where the optical depth $\tau>>1$ and $(v/c)\tau>>1$, this radiative flux can dominate the diffusion flux, emission or absorption. For an alternative mixed frame formulation, see \cite{Krumholz_et_al_2007}.
In the low-mass star domain, the main focus of this work, we do not expect to encounter dynamic diffusion situations.

\subsubsection{The flux-limited diffusion approximation\label{sec_FLD}}

As mentioned earlier, we need a closure relation to solve the moment equations coupled to the hydrodynamics (closed by the perfect gas relation), and such a relation is of prime importance. Many possible choices for the closure relation exist. Among these models, the FLD approximation is one of the simplest ones and uses moment models of radiation transport \citep{Minerbo_1978JQSRT,Levermore_Pomraning_1981ApJ}.

Under the flux-limited diffusion approximation,
the  radiative  flux  is  expressed  directly as  a  function  of  the
radiative  energy  and is  proportional and  colinear to  the radiative
energy  gradient (Fick's law).  Under the grey approximation, we have
\begin{equation}
\textbf{F}_\mathrm{r} = -\frac{c\lambda}{\sigma_\mathrm{R}}\nabla E_\mathrm{r} \label{diff2},
\end{equation}
where  $\lambda  =  \lambda  (R)$  is  the  flux  limiter, and  $R=|\nabla  E_\mathrm{r} |/(\sigma_\mathrm{R} E_\mathrm{r})$.
In this study, we retain
the flux limiter that has been
derived by \cite{Minerbo_1978JQSRT}, assuming the intensity as a piecewise
linear function of solid angle
\begin{equation}
\lambda=\left\{\begin{array}{ccc}
2/(3+\sqrt{9 + 12 R^ 2}) &  \mathrm{if } &    0\le R \le 3/2,\\
(1+ R + \sqrt{1 + 2 R})^{-1} & \mathrm{if } &  3/2 <  R \le \infty.
\end{array}\right.
\end{equation}
The  flux  limiter  has  the property  that  $\lambda
\rightarrow 1/3$  in optically thick regions  and $\lambda \rightarrow
1/R$  in optically  thin regions.   We  recover the  proper value  for
diffusion               in            the    optically               thick
 regime, $\mathrm{F}=-c/(3\sigma_\mathrm{R})\nabla  E_\mathrm{r}$, and
the  flux is  limited  to $cE_\mathrm{r}$  in the optically thin  regime. Under the FLD approximation, the radiative transfer equation is then replaced by a unique diffusion equation on the radiative energy
\begin{equation}
\label{nrjrad} 
\frac{\partial E_\mathrm{r}}{\partial t} - \nabla \cdot\left(\frac{c\lambda}{\sigma_\mathrm{R}} \nabla E_\mathrm{r}\right) = 
\sigma_\mathrm{P}(4\pi B -cE_\mathrm{r}).
\end{equation}


\section{A multidimensional radiation hydrodynamics solver for {\ttfamily RAMSES}\label{FLD_RAMSES}}

\subsection{The AMR {\ttfamily RAMSES} code}

We use the {\ttfamily RAMSES} code \citep{teyssier-2002}, which integrates the equations of ideal magnetohydrodynamics \citep{Fromang_2006,Teyssier_2006} using a second-order Godunov finite volume scheme. The MHD equations are integrated using a MUSCL predictor-corrector scheme, originally presented in  \cite{vanLeer_1979}. Fluxes at the cell interface are estimated with an approximated Riemann solver (Lax-Friedrich, HLL, HLLD, etc...).
For its AMR grid, {\ttfamily RAMSES} is based on a `tree-based'' AMR structure, the refinement is made on a cell-by-cell basis. Various refinements can be used (fluid variable gradients, instability wavelength, etc...). 

The AMR code {\ttfamily RAMSES} has often been used for star-formation purposes \citep{Hennebelle_Fromang_2008,Hennebelle_Teyssier_2008, Hennebelle_Ciardi_2009,Commercon_2010L}. \cite{Commercon_2008} have thoroughly and successfully compared its results with standard SPH, showing a good agreement between the methods.

\subsection{The Eulerian approach and properties of conservation laws}

In Eulerian hydrodynamics, the mesh is fixed and gas density, velocity, and internal energy are primary variables. Eulerian methods fall into two groups: finite difference methods \citep[e.g.  the {\ttfamily ZEUS} code, ][]{Stone_Norman_92,Turner_stone_01} and finite volume methods \citep[e.g. the {\ttfamily RAMSES} code, ][]{teyssier-2002}.  In the first group, flow variables are conceived as being samples at certain points in space and time. Partial derivatives are then computed from these sampled values and follow  Euler equations. In the finite volume approach, flow variables correspond to average values over a finite volume - the cell - and obey the conservation laws in the integral form. Their evolution is determined through Godunov methods by calculating the flux of every conserved quantity across  each cell interface.\\ 

For an inviscid, compressible flow, the Euler equations in their conservative form read
\begin{equation}
\left\{
\begin{array}{ccccl}
\partial_t \rho & + & \nabla \left[\rho\textbf{u} \right] & = & 0 \\
\partial_t \rho \textbf{u} & + & \nabla \left[\rho \textbf{u}\otimes \textbf{u} + P \mathbb{I} \right]& =& 0 \\
\partial_t E & + & \nabla \left[\textbf{u}\left( E + P \right)\right] &= & 0 \\
 \end{array}
\right.
\end{equation}
This system can be written in the general hyperbolic conservative  form

\begin{equation}
\frac{\partial \mathbb{U}}{\partial t} + \nabla.\mathbb{F}(\mathbb{U})=0,
\label{syst}
\end{equation}
where the vector $\mathbb{U}=(\rho, \rho \textbf{u},E)$ contains conservative variables, and the flux vector $\mathbb{F} (\mathbb{U})=(\rho \textbf{u}, \rho \textbf{u} \otimes \textbf{u} + P \mathbb{I}, \textbf{u}\left(\mathrm{E} + P\right))$ is a linear function of    $\mathbb{U}$.

 In this paper, we use the second-order Godunov method, but applied to the modified Euler equation system under the FLD approximation.

\subsection{The conservative radiation hydrodynamics scheme}

Let us rewrite the grey RHD equations under the FLD approximation within the comoving frame, taking into account the gravity terms
 
 \begin{equation}
\left\{
\begin{array}{llll}
\partial_t \rho + \nabla \left[\rho\textbf{u} \right] & = & 0 \\
\partial_t \rho \textbf{u} + \nabla \left[\rho \textbf{u}\otimes \textbf{u} + P \mathbb{I} \right]& =& -\rho\nabla\Phi - \lambda\nabla E_\mathrm{r} \\
\partial_t E_\mathrm{T} + \nabla \left[\textbf{u}\left( E_\mathrm{T} + P_\mathrm{} \right)\right] &= &-\rho\textbf{u}\cdot\nabla \Phi - \mathbb{P}_\mathrm{r}\nabla:\textbf{u}  - \lambda \textbf{u} \nabla E_\mathrm{r} \\
 & & +  \nabla \cdot\left(\frac{c\lambda}{\rho \kappa_\mathrm{R}} \nabla E_\mathrm{r}\right) \\
\partial_t E_\mathrm{r} + \nabla \left[\textbf{u}E_\mathrm{r}\right]
&=& 
- \mathbb{P}_\mathrm{r}\nabla:\textbf{u}  +  \nabla \cdot\left(\frac{c\lambda}{\rho \kappa_\mathrm{R}} \nabla E_\mathrm{r}\right) \\
 & &  + \kappa_\mathrm{P}\rho c(a_\mathrm{R}T^4 - E_\mathrm{r})
\end{array}
\right.
\end{equation}
Note that we rewrite the opacity $\sigma_\mathrm{i}$ as $\kappa_\mathrm{i}\rho$. The dimension of $\kappa_\mathrm{i}$  is  cm$^2$ g$^{-1}$.
\\

The basic idea is to build a solver for a radiative fluid, with an additional pressure owing to the radiation field: the radiative pressure. Following the Euler equations in their conservative form, the new conservative quantities are density $\rho$, momentum $\rho \textbf{u}$, total energy $E_\mathrm{T}$ of the fluid (gas + photon) per unit volume, i.e. $\rho
\epsilon  + \rho u^2/2  + E_\mathrm{r}$. Primitive hydrodynamical variables do not change for the fluid, but we add a fourth equation for the radiative energy.

In order to facilitate these equations in {\ttfamily RAMSES} and to minimize the number of  changes with the pure hydrodynamical version, we decompose each term where the flux limiter $\lambda$  appears as follows: $\lambda  = 1/3  +  (\lambda -  1/3)$. We thus  distinguish a diffusive part  (Eddington approximation,
$\mathbb{P}_\mathrm{r}       =1/3E_\mathrm{r}\mathbb{I}$) and a correction part. The computation of predicted states and fluxes in the MUSCL scheme is  made under the Eddington approximation, which is then corrected in an additional corrective step. The RHD equations can be rewritten as
{\small
\begin{equation}
\left\{
\begin{array}{llll}
\partial_t  \rho +  \nabla \left[\rho\textbf{u}  \right]  & =  & 0  \\
\partial_t  \rho  \textbf{u}  +  \nabla  \left[\rho  \textbf{u}\otimes
\textbf{u}   +   (P   +1/3   E_\mathrm{r}) \mathbb{I}      \right]&   =&
-\rho\nabla\Phi - (\lambda - 1/3)\nabla E_\mathrm{r} \\ 
\partial_t E_\mathrm{T} +
\nabla \left[\textbf{u}\left( E_\mathrm{T}  + P  + 1/3 E_\mathrm{r}\right)\right] &=
&-\rho\textbf{u}\cdot\nabla  \Phi  \\
 & & -  (\lambda-1/3) (\textbf{u}  \nabla
E_\mathrm{r} +E_\mathrm{r}\nabla:\textbf{u}) \\
& & +     \nabla
\cdot\left(\frac{c\lambda}{\rho        \kappa_\mathrm{R}}       \nabla
E_\mathrm{r}\right)\\
\partial_t  E_\mathrm{r} +  \nabla \left[\textbf{u}E_\mathrm{r}\right]
&=&          -\mathbb{P}_\mathrm{r}\nabla:\textbf{u}          +
\kappa_\mathrm{P}\rho c(a_\mathrm{R}T^4 - E_\mathrm{r})  \\
& &    +     \nabla
\cdot\left(\frac{c\lambda}{\rho        \kappa_\mathrm{R}}       \nabla
E_\mathrm{r}\right)
\end{array}
\right. .
\label{sys_ramses}
\end{equation}
}
The new system $\partial_t\mathbb{U}+\nabla\mathbb{F}(\mathbb{U})=S(\mathbb{U})$  is composed of the modified hyperbolic left hand side (LHS) and the right hand side (RHS) source, corrective and coupling terms $S(\mathbb{U})=S_\mathrm{exp}+S_\mathrm{imp}$. The hyperbolic system, as well as  the source and corrective terms $S_\mathrm{exp}$, are integrated in time with an explicit scheme. The modified RHD hyperbolic system reads
\begin{equation}
\begin{array}{ccc}
\mathbb{U} = 
\left[
\begin{array}{c}
\rho \\ \rho\textbf{u}\\E_\mathrm{T}\\E_\mathrm{r}
\end{array}
\right]
& , &
\mathbb{F}(\mathbb{U}) = 
\left[
\begin{array}{c}
\rho\textbf{u} \\ \rho  \textbf{u}\otimes
\textbf{u}   +   (P   +1/3   E_\mathrm{r}) \mathbb{I}\\
\textbf{u}\left( E_\mathrm{T}  + P  + 1/3 E_\mathrm{r}\right)\\\textbf{u}E_\mathrm{r}
\end{array}
\right]
\end{array}.
\label{RHD}
\end{equation}
This system is used in the predictor-corrector MUSCL temporal integration. To predict states, we consider the worst case, where the radiative pressure is the greatest ($1/3E_\mathrm{r}$). For the conservative update (corrector step) we consider the LHS of system (\ref{sys_ramses}). 
The associated eigenvalues corresponding to the three waves are
\begin{equation}
  \lambda_i = \left\{
\begin{array}{l}
u - \sqrt{\frac{\gamma P}{\rho} + \frac{4 E_\mathrm{r}}{9\rho}}\\
u \\
u + \sqrt{\frac{\gamma P}{\rho} + \frac{4 E_\mathrm{r}}{9\rho}}\\
\end{array}
\right. .
\end{equation}
Radiative pressure enlarges the span of solutions, since wave speeds are faster. Once again, with the Eddington approximation, we build the system for the case where the radiative pressure would be the greatest. Therefore, the waves propagate at a speed that is within the wave extrema. 

The next step consists in correcting errors due to the Eddington approximation by integrating source terms $S_\mathrm{ne}$ 

\begin{equation}
S_\mathrm{ne}=\left(
\begin{array}{c}
0 \\
 - (\lambda - 1/3)\nabla E_\mathrm{r} \\
 - (\lambda - 1/3)(\textbf{u}\nabla E_\mathrm{r}  + E_\mathrm{r}\nabla:\textbf{u})\\
\mathbb{P}_\mathrm{r}\nabla:\textbf{u}
\end{array}
\right) .
\end{equation}
We here consider an isotropic radiative pressure tensor $\mathbb{P}_\mathrm{r}=\lambda E_\mathrm{r}\mathbb{I}$. Other authors considered extensions to this closure relation using the \cite{Levermore_1984}  FLD theory  \citep{Turner_stone_01,Krumholz_et_al_2007}.

To ensure the stability of the explicit step, the Courant-Friedrich-Levy (CFL) stability condition used to estimate the timestep  also  takes into account the radiative pressure. The updated CFL condition is simply
\begin{equation}
\Delta t \leq C_\mathrm{CFL}\frac{\Delta x}{u+\sqrt{\frac{\gamma P}{\rho}+\frac{4E_\mathrm{r}}{9\rho}}}.
\end{equation}

\subsection{The implicit radiative scheme\label{imp_part}}
The most demanding step in our time-splitting scheme is to deal with the diffusion term $\nabla\cdot\left(\frac{c\lambda}{\rho        \kappa_\mathrm{R}}       \nabla E_\mathrm{r}\right)$ and the coupling term $\kappa_\mathrm{P}\rho c(a_\mathrm{R}T^4 - E_\mathrm{r})  $,  which corresponds to $S_\mathrm{imp}$. This update has to be made with an {\bf implicit} scheme, because the time scales of these processes are much shorter than those of pure hydrodynamical processes. 
 Two coupled equations are integrated implicitly
\begin{equation}
\left\{
\begin{array}{lll}
\partial_t\rho\epsilon & = & - \kappa_\mathrm{P}\rho \mathrm{c}(a_\mathrm{R}T^4 - E_\mathrm{r})\\
\partial_t E_\mathrm{r} - \nabla \frac{\mathrm{c}\lambda}{\kappa_\mathrm{R}\rho}\nabla E_\mathrm{r} & = & + \kappa_\mathrm{P}\rho \mathrm{c}(a_\mathrm{R}T^4 - E_\mathrm{r})
\end{array}
\right. ,
\end{equation}
which give the implicit scheme on a uniform grid{\footnote{Index $n$ and $n+1$ are used for variables before and after the implicit update. Outputs of the explicit hydrodynamics scheme supply variables with index $n$. They do not match the  variables at time $t^n$ and $t^{n+1}$.}

\begin{equation}
\left\{\begin{aligned}
\frac{C_v T^{n+1} - C_v T^{n}}{\Delta t} & = & - \kappa_\mathrm{P}^n\rho^n \mathrm{c}(a_\mathrm{R}(T^{n+1})^4 - E_\mathrm{r}^{n+1}) \\ 
\frac{E_\mathrm{r}^{n+1} - E_\mathrm{r}^{n}}{\Delta t}  - \nabla \frac{\mathrm{c}\lambda^n}{\kappa_\mathrm{R}^n\rho^n}\nabla E_\mathrm{r}^{n+1} & = & + \kappa_\mathrm{P}^n\rho^n \mathrm{c}(a_\mathrm{R}(T^{n+1})^4 - E_\mathrm{r}^{n+1})
\end{aligned}
\right.  ,
\label{imp_schema}
\end{equation}

where $\rho \epsilon = C_v T$. The nonlinear term $(T^{n+1})^4$ makes this scheme difficult to invert. Yet it is much easier to solve implicitly a linear system. Assuming that changes of temperature are small within a time step, we can write
\begin{equation}
(T^{n+1})^4  = (T^{n})^4\left(1 + \frac{(T^{n+1}-T^{n})}{T^{n}}\right)^4 \approx 4 (T^{n})^3T^{n+1} 
-3(T^{n})^4,
\label{lin_tp}
\end{equation}
Eventually, with (\ref{imp_schema}a), we obtain $T^{n+1}_i$ as a function of $T^{n}_i$ and $E_{\mathrm{r},i}^{n+1} $. Then  $T^{n+1}_i$ can be directly injected in the radiative energy equation (\ref{imp_schema}b), and  $E_{\mathrm{r},i}^{n+1}$ is finally expressed  as a function of  $E_{\mathrm{r},i+1}^{n+1}$, $E_{\mathrm{r},i-1}^{n+1}$, $E_{\mathrm{r},i}^{n}$ and $T^{n}_i$. The implicit scheme for the radiative energy in a cell of volume $V_i$ in the $x-$direction becomes

\begin{eqnarray}
(E^{n+1}_{r,i} - E^{n}_{r,i} )V_i
& - & c \Delta t \left( \frac{ \lambda}{\kappa_{\mathrm{R}\rho}}\right)_{i+1/2} S_{i+1/2} \frac{E^{n+1}_{r,i+1} - E^{n+1}_{r,i} }{\Delta x_{i+1/2}}\nonumber \\
& + &  c \Delta t \left(\frac{\lambda}{\kappa_{\mathrm{R}\rho}}\right)_{i-1/2} S_{i-1/2} \frac{E^{n+1}_{r,i} - E^{n+1}_{r,i-1} }{\Delta x_{i-1/2}}   \label{imp_scheme_1}\\ 
 & = & c \Delta t  \kappa_{\mathrm{P},i}^n  \rho^n_i\left( 4a_\mathrm{R} (T_i^n)^3 T_i^{n+1} - 3 a_\mathrm{R} (T_i^n)^4 - E^{n+1}_{r,i}\right)V_i.
\nonumber
\end{eqnarray}
The gas temperature within a cell is simply given by
\begin{equation}
T^{n+1}_i=\frac{3a_\mathrm{R}\kappa_{\mathrm{P},i}^n c\Delta t \left(T^n_i\right)^4 + C_vT^n_i+ \kappa_{\mathrm{P},i}^n c\Delta t E_{r,i}^{n+1}}{C_v + 4a_\mathrm{R}\kappa_{\mathrm{P},i}^n c\Delta t \left(T^n_i\right)^3 }.
\end{equation}
We compute the Planck and Rosseland opacities and the flux limiter with a gas temperature value given before the implicit update (with index $n$)  to preserve the linearity of the solver. 

\subsection{Implicit scheme integration with the conjugate gradient algorithm}
Equation (\ref{imp_scheme_1}) is solved on a full grid made of $N$ cells. It results in a system of $N$ linear equations, which can be written as a linear system of equations
\begin{equation}
A{\bf x}=b,
\end{equation}
where ${\bf x}$ is a vector containing radiative energy values.
The conjugate gradients (CG) method is one of the most popular non-stationary iterative methods for solving large symmetric systems of linear equations $A{\bf x}=b$.
The CG method can be used if the matrix $A$ to be inverted is square,  symmetric, and positive-definite. The CG is memory-efficient (no matrix storage) and runs quickly with sparse matrices. For a $N\times N$ matrix, the CG converges in less than $N$ iterations. Basically, the CG method is a steepest-gradient-descent method in which descent directions are updated at each iteration. 
Another advantage is that the CG method can be run easily on parallel machines.

To improve convergence of the CG or even to insure convergence if one deals with an ill-conditioned matrix $A$, we use a preconditioning matrix $M$ that approximates $A$. $M$ is also assumed to be  symmetric and positive definite. 
In this work, we use a simple diagonal  preconditioning matrix, which retains only the inverse of $A$ diagonal elements. The convergence of the CG algorithm is estimated following two criteria: estimation of the norm $L^2$ (criterion $|| r^{(j)} ||/|| r^{(0)}|| < \epsilon $) or estimation of the norm $L^\infty$ (maximum residual value $\mathrm{max}\{r^{(j)}\}/\mathrm{max}\{r^{(0)}\}<\epsilon $). Values of $\epsilon$ typically range from $10^{-8}$ to $10^{-3}$. 
 In appendix \label{STS} we present an alternative method to the conjugate gradient, the Super-Time Stepping method, which can be used efficiently on uniform grids or in some particular cases. \\

\subsection{Comparison to other schemes}

 Other RHD solvers based on the FLD approximation have been designed in grid-based codes. Among them, the {\ttfamily ZEUS} and {\ttfamily ORION} implementations are the most widely used and discussed. Compared to {\ttfamily ZEUS} \citep{Stone_Norman_92,Turner_stone_01,Hayes_2006}, our method is fundamentally different, although they also use the comoving frame to estimate the radiative quantities. {\ttfamily ZEUS} code is based on a finite difference scheme, using artificial viscosity and regular grids. Its non-conservative formulation can lead to spurious wave propagation when the resolution is not high enough, or if the radiative pressure dominates the characteristic velocity (the classical Burgers equation problem). All radiative terms such as the radiation transport and the radiative pressure work are integrated implicitly in {\ttfamily ZEUS}. The implicit scheme is based on a Newton-Raphson method, using GMRES or LU algorithms for the matrix inversion. 

{\ttfamily ORION} is a patched-based AMR code, which is less flexible than the tree-based AMR \citep{Krumholz_07}. \cite{Krumholz_07} implemented the mixed-frame RHD equations using a multi-grid, multi-timestep method to solve the implicit scheme for the radiation module  \citep{Howell_Greenough_03}. The hydrodynamics part of {\ttfamily ORION} uses a second-order conservative Godunov scheme,  with approximate Riemann solvers and very little artificial viscosity to treat shocks and discontinuities. Using the same idea as in this study, the diffusion and matter-radiation coupling terms are integrated implicitly, while the radiative force and radiative pressure work are integrated explicitly. Contrary to our work,  \cite{Krumholz_07} do not take into account the radiative pressure in the flux estimate at the cell interface for the conservative update, which could also lead to an inaccurate wave speed propagation in radiation-pressure dominated regions.

\subsection{Implicit scheme on an AMR grid\label{amr_part}}

For studies involving large dynamical ranges, such as star formation, it is necessary to extend our implicit scheme to AMR grids. The difficulty is to compute the correct fluxes and gradients at the interfaces between two cells. We need to carefully consider the energy balance on a given volume.  
Energy balances are performed on volumes overlapping two cells, which depends on whether the mesh is refined or not. If one considers the face of a cell on a level $\ell$; three connecting configurations with other cells are possible (see Fig. \ref{ex_AMR}):
\begin{itemize}
\item Configuration 1: the neighbouring cell is at the same level $\ell$: cells 1 and 3,
\item Configuration 2: the neighbouring cell is at level $\ell -1$: cells 1 and i,
\item Configuration 3: 2 neighbouring cells exist at level $\ell +1$: cell i with cells 1 and 2.
\end{itemize}

Last but not least, the diffusion routine is called only once per coarse step (no multiple timestepping) and scans the full grid, from the finer level to the coarse level. In order to optimise matrix-vectors products,
we choose to avoid dealing with configuration 3. Hence, when cells at level $\ell+1$ are monitored, values for cells at level $\ell$ are updated. Configurations 2 and 3 are then performed at the same time. Depending on the configuration, gradients and flux estimates are different. In the following, we will focus on cell 1, of size $\Delta x \times \Delta x $.

\begin{figure} [thb]
\centering
\includegraphics{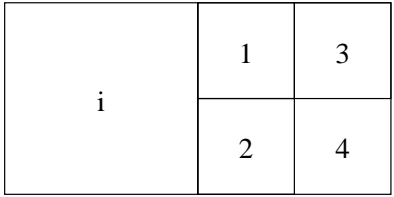}
\caption{Example of AMR grid configuration}
\label{ex_AMR}
\end{figure}

\subsubsection{Gradient estimate}
Gradients $\nabla E_r$ are estimated between the two neighbouring cells centre

\begin{equation*}
\begin{aligned}
-\hspace{3pt}\mathrm{ Configuration} \hspace{3pt}1& :  (\nabla E_r)_{1,3} = \frac{E_{r,1}-E_{r,3}}{\Delta x}\hspace{100pt} \\
-\hspace{3pt}\mathrm{ Configuration} \hspace{3pt}2& : (\nabla E_r)_{1,i} = \frac{E_{r,1}-E_{r,i}}{3 \Delta x/2}
\end{aligned}
\end{equation*}

\subsubsection{Flux estimate}

Let $S^\ell$ be the surface of the interface of a cell at level $\ell$ and $F^\ell_{i,j}$ the flux across this surface between two cells $i$ and $j$. The energy rate $F\times S$ that is exchanged at this interface is

\begin{equation*}
\begin{aligned}
-\hspace{3pt}\mathrm{ Configuration} \hspace{3pt}1& :  F^\ell_{1,3}\times S^\ell  =  \frac{E_{r,1}-E_{r,3}}{\Delta x} \times S^\ell \hspace{100pt}\\
-\hspace{3pt}\mathrm{ Configuration} \hspace{3pt}2& : F^\ell_{1,i}\times S^\ell =  \frac{E_{r,1}-E_{r,i}}{3\Delta x/2} \times S^\ell \\
-\hspace{3pt}\mathrm{ Configuration} \hspace{3pt}3& : F^\ell_{i,1,2}\times S^{\ell-1} = F^\ell_{i,1}\times S^\ell + F^\ell_{i,2}\times S^\ell 
\end{aligned}
\end{equation*}

Because access to neighbouring finer cells is not straightforward, we see from configuration 3 all the interest of updating quantities at level $\ell -1$ when scanning grid at level $\ell$.

\subsection{Limits of the methods}

The first drawback is the use of the FLD approximation that implies isotropy of the radiation field. Anisotropies in the transparent regime are not well processed with the FLD, contrary to more accurate models like M1\citep{Gonzalez_2007} or VETF \citep{Hayes_Norman_03}. A second limitation comes from the grey opacity assumption, which could limit the accretion on the protostars \citep[see][ and references therein]{Zinnecker_Yorke_2007}.  In high-mass star formation, \cite{Yorke_Sonnhalter_2002} show that using a frequency-dependent radiative transfer model enhances the flashlight effect and helps to accrete more mass onto the central protostar.

From a technical point of view, our method works only for unique time stepping, i.e. all levels evolve with the same time step. We do not take advantage of the multiple time stepping possibility. As a compromise, we investigate the possibility to evolve finer levels
with their own time steps and perform a diffusion-coupling step every 2, 4 or more finer time steps. As a result, we find that performing the diffusion step only every 2 or 4 fine time steps gives correct results. The frequency of the implicit solver calls is left to the user convenience, by use of the mutli-time stepping of {\ttfamily RAMSES}. For instance, for a grid of levels ranging form level $\ell_\mathrm{min}$ to $\ell_\mathrm{max}$, a unique timestep can be used for levels ranging form level $\ell_\mathrm{min}$ to $\ell_i$. In that case, only the levels finer that $\ell_i$ will use not updated radiative quantities in the Godunov solver. 
 A future development would be to use a multigrid solver or preconditioner for parabolic equations \citep{Howell_Greenough_03}.

Another difficulty comes from the residual norm and scalar  estimates in the CG algorithm. For a large grid with a large number of cells, the dot product can be dominated by round-off errors, owing to estimates close to machine precision. This becomes even worse in parallel calculations. The usual MPI function MPI\_SUM fails with a large number of processors, the results of any sum  becoming a function of number of processors. This dramatically  affects  the number of iterations. We implemented a new MPI function that performs summation in  double-double precision following \cite{Yu_Ding_MPI_SUMDD}, using the  \cite{Knuth_97} method. 

Eventually, our method involves only immediate neighbouring cells, whatever their refinement level. As a consequence, our method is only first order at the border between levels. This could give rise to  a loss of accuracy in diffusion problems, because gradient estimates are not second-order accurate when neighbouring cells are at finer levels \citep[see configuration 3, ][]{Popinet_03}. However, the tests we performed ascertain that the method is still {\it roughly} second-order accurate. The errors are only confined to surfaces much smaller than the total volume.

\section{Radiation hydrodynamics solver tests}

\subsection{1D test: linear diffusion\label{1D_diff_lin}}

We only consider the radiative energy diffusion equation in this test, without either hydrodynamics or coupling with the gas. 
The equation to integrate is simply	
\begin{equation}
\frac{\partial E_\mathrm{r}}{\partial t} - \nabla \cdot\left(\frac{c}{3 \rho \kappa_\mathrm{R}} \nabla E_\mathrm{r}\right) = 0.
\end{equation}
Consider a box of length L=1. The initial radiative energy corresponds to a delta function, namely it is equal to 1 everywhere in the box, except at the centre, where it  equals $E_\mathrm{r,L/2}\Delta x=E_0=1\times 10^5$. To simplify, we choose $\rho\kappa_\mathrm{R}=1$ and a constant time step. We apply Von Neumann boundary conditions, i.e. zero-gradient. The analytical solution in a $p$-dimensional problem is given by
\begin{equation}
E_{\mathrm{r},a}(x,t)=\frac{E_0}{2^{p}(\pi \chi t)^{p/2}}e^{-\left(\frac{x^2}{4\chi t}\right)},
\end{equation}
where $\chi=c/(3\rho \kappa_\mathrm{R})$.

\begin{figure}[t]
  \centering
 \resizebox{\hsize}{!}{ \includegraphics{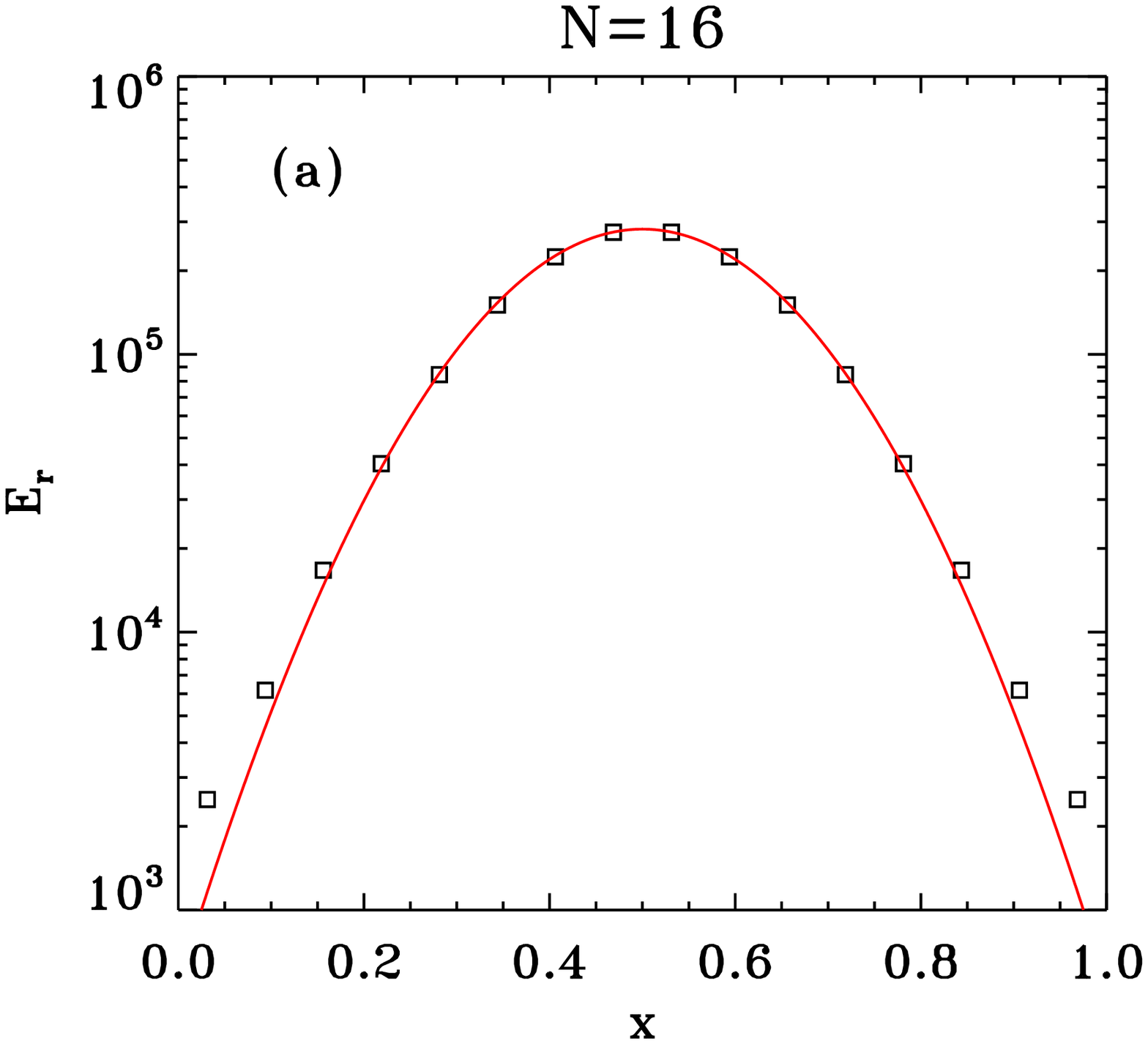}}
 \resizebox{\hsize}{!}{ \includegraphics{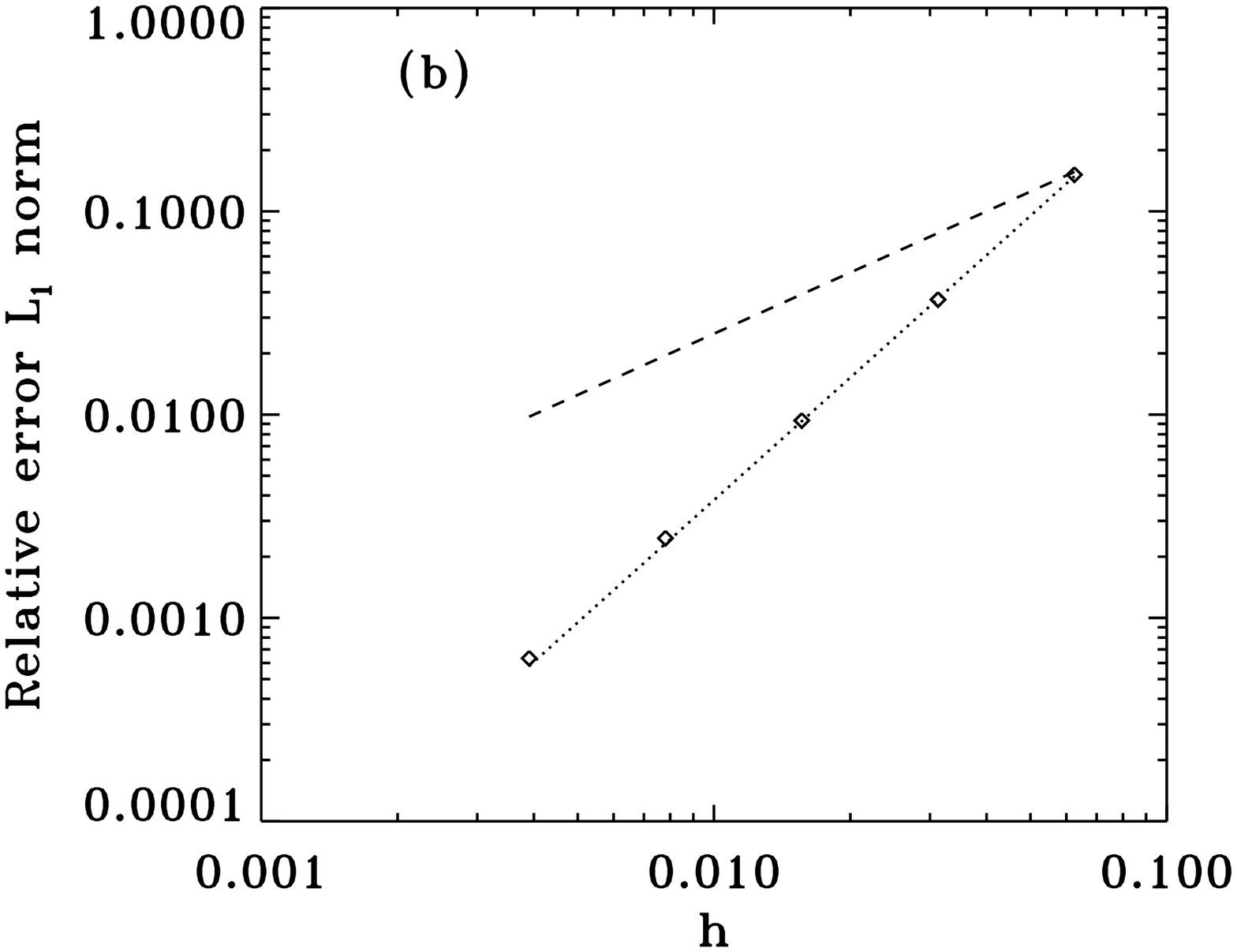}}
  \caption{(a): Comparison  between numerical solution  (squares)  and analytical solution (red line) at time t=$1\times 10^{-12}$ for the calculations with 16 cells. 
  (b): L$_1$ norm of the error as a function of $h=1/\Delta x$. The dotted line shows the evolution of the error as a function of $h^2$ and the dashed line the evolution of the error as a function of $h$.} 
\label{diff_lin}
\end{figure}
 
\begin{table}[htb]
\caption{CPU time, total number of time steps, and number of iterations per time step for various numbers of cells N.}
\center
\begin{tabular}{c c c c}
\hline\hline
 N & CPU time (s) & $N_\mathrm{\Delta t}$ &$ N_\mathrm{iter}/N_{\Delta t}$\\
\hline
   32            &   3.5     &   51    & 9.9\\
   64            &   7.1     &  102  & 10.4\\ 
  128           &  14.36  &   205  &  10.6\\ 
  256           &  30.85  &   409  &  11.8 \\ 
\hline
\end{tabular} 
\label{diff_lin_sum}
\end{table}

Figure \ref{diff_lin}(a) shows results at time $t=1\times 10^{-12}$ for a resolution of  $N=16$ cells. The numerical solution is very close to the analytical one, even with this small number of cells. In Fig.  \ref{diff_lin}(b) we show the evolution of the L$_1$ norm of the relative error as a function of $h=1/\Delta x$. The L$_1$ norm is estimated as
\begin{equation}
\mathrm{L}_1 = \sqrt{\frac{\sum_1^N{|E_{\mathrm{r},i} - E_{\mathrm{r},a}(x_i,t) |\Delta x_i}}{\sum^N_1 E_{\mathrm{r},a}(x_i,t)\Delta x_i}},
\end{equation}
where $E_{\mathrm{r},i}$ is the numerical value of  the radiative energy  at position $x_i$ and time $t$, and $E_{\mathrm{r},a}(x_i,t)$ the corresponding analytic value. The error clearly grows as $h^2$ (dotted line), which indicates that our method is second-order accurate.

In Table \ref{diff_lin_sum} we report the CPU time, the total number of iterations, and the number of time steps for various numerical resolutions. At low resolution, the number of time steps increases linearly with the number of cells, as well as the CPU time. The number of iterations per time step is constant, i.e. the convergence of CG does not depend on the dimension of the problem, but on the nature of the problem.

\subsection{1D test: nonlinear diffusion}


In this second test, we consider an initial discontinuity in a box with different initial radiative energy states:  $E_\mathrm{r}=4$ on the left and  $E_\mathrm{r}=0.5$  on the right. We apply Von Neumann boundary conditions. We integrate the same equation as in the previous test, but with a Rosseland opacity as a nonlinear function of the radiative energy, i.e. $\rho\kappa_\mathrm{R}=1\times 10^{11}E_\mathrm{r}^{-1.5}$. Last, we allow refinement with a criterion based on the radiative energy gradient. In each region where $\nabla E_\mathrm{r}/E_\mathrm{r} > 3$ \%, the grid is refined. 

 Figure \ref{diff_non_lin}({\it a}) shows the radiative energy profiles at time $t= 1.4\times 10^{-2}$ for calculations run with a coarse grid of 16 cells and a maximum effective resolution of 512 cells (squares), and for calculations run with 2048 cells, taken to be the "exact" solution (red curve). Because of the nonlinear opacity, the diffusion is more efficient in the high-energy region. The mean opacity at cell interface is computed using an arithmetic average, which is more adapted for the case of nonlinear opacity. The levels are finer (higher resolution) in high-radiative energy gradient regions. Note that we checked that we obtained similar results in a 2D plane parallel case and in a 2D case with an initial step function that maked an angle $\pi/4$ with the computational box axis. This validates our routine in the $x$ and $y$ directions. 

In Fig.  \ref{diff_non_lin}({\it b}) we show the evolution of the L$_1$ norm of the error as a function of the mesh spacing. The uniform grid points (diamonds) correspond to a calculations run with a number of cells ranging from 16 to 512 (i.e. $\ell=4$ to $\ell=9$) . When AMR is used (squares), the error is plotted as function of the minimum grid spacing, corresponding to effective resolutions ranging from 32 to 512  cells. The coarse level remains unchanged,  $\ell_\mathrm{min}=4$ (i.e. 16 coarse cells).
 The advantage of the AMR is clear, the error remains identical compared to the uniform grid case, but  the number of cells is greatly reduced (25 cells with  $\ell_\mathrm{max}=5$, 38 cells with  $\ell_\mathrm{max}=6$, 59 cells with  $\ell_\mathrm{max}=7$, 83 cells with  $\ell_\mathrm{max}=8$ and 110 cells with  $\ell_\mathrm{max}=9$). The AMR implementation works well (second-order accuracy), and does not suffer from the fact that our scheme is only first order in space at the level interface, which validates our scheme used to estimate the gradients at the cell interface in Sect. \ref{amr_part}.
 Finally, as in the previous test, the error increases with $h^2$, even if the diffusion problem is nonlinear for radiation.

\begin{figure}[htb]
  \centering
   \resizebox{\hsize}{!}{ \includegraphics{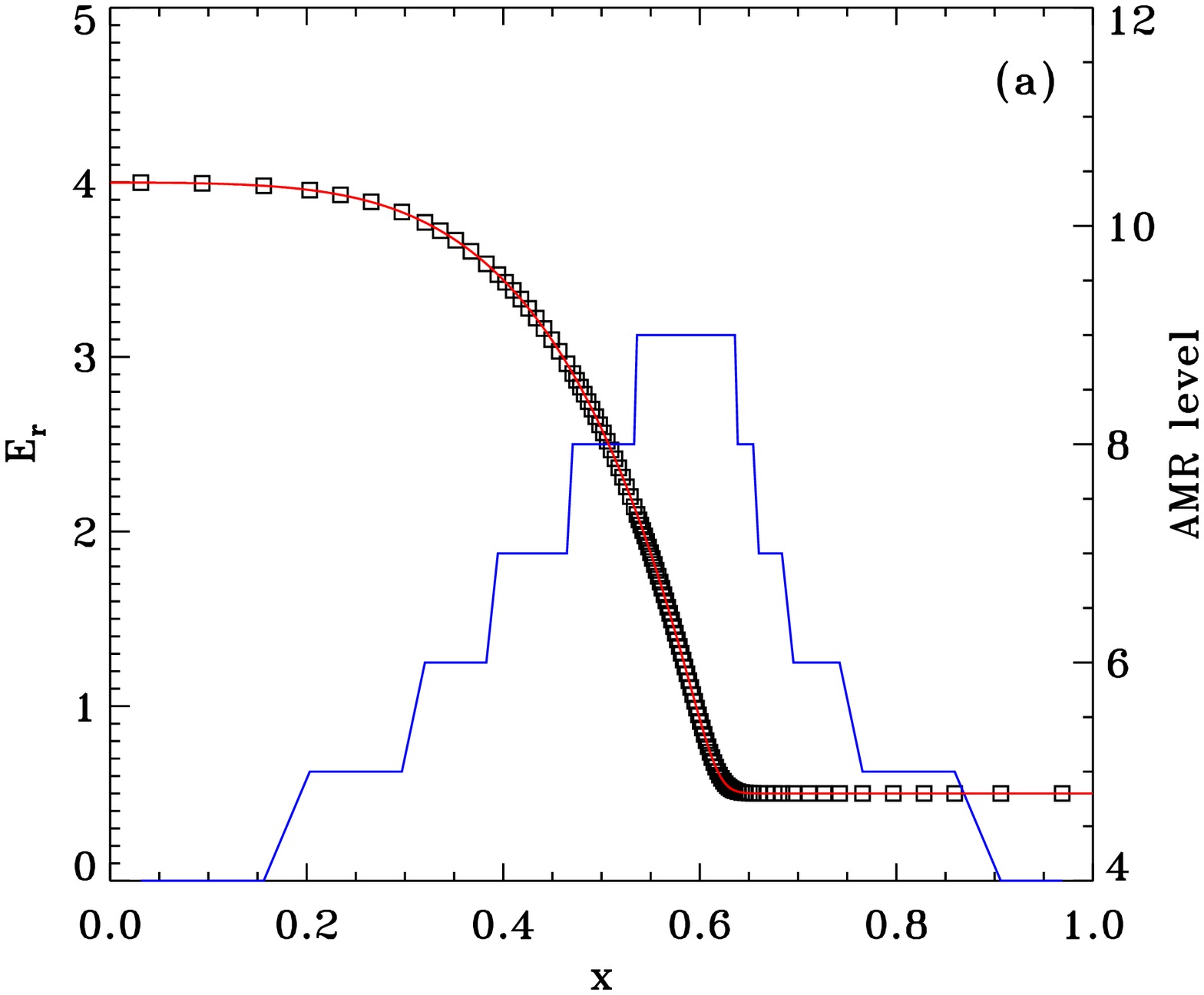}}
   \resizebox{\hsize}{!}{ \includegraphics{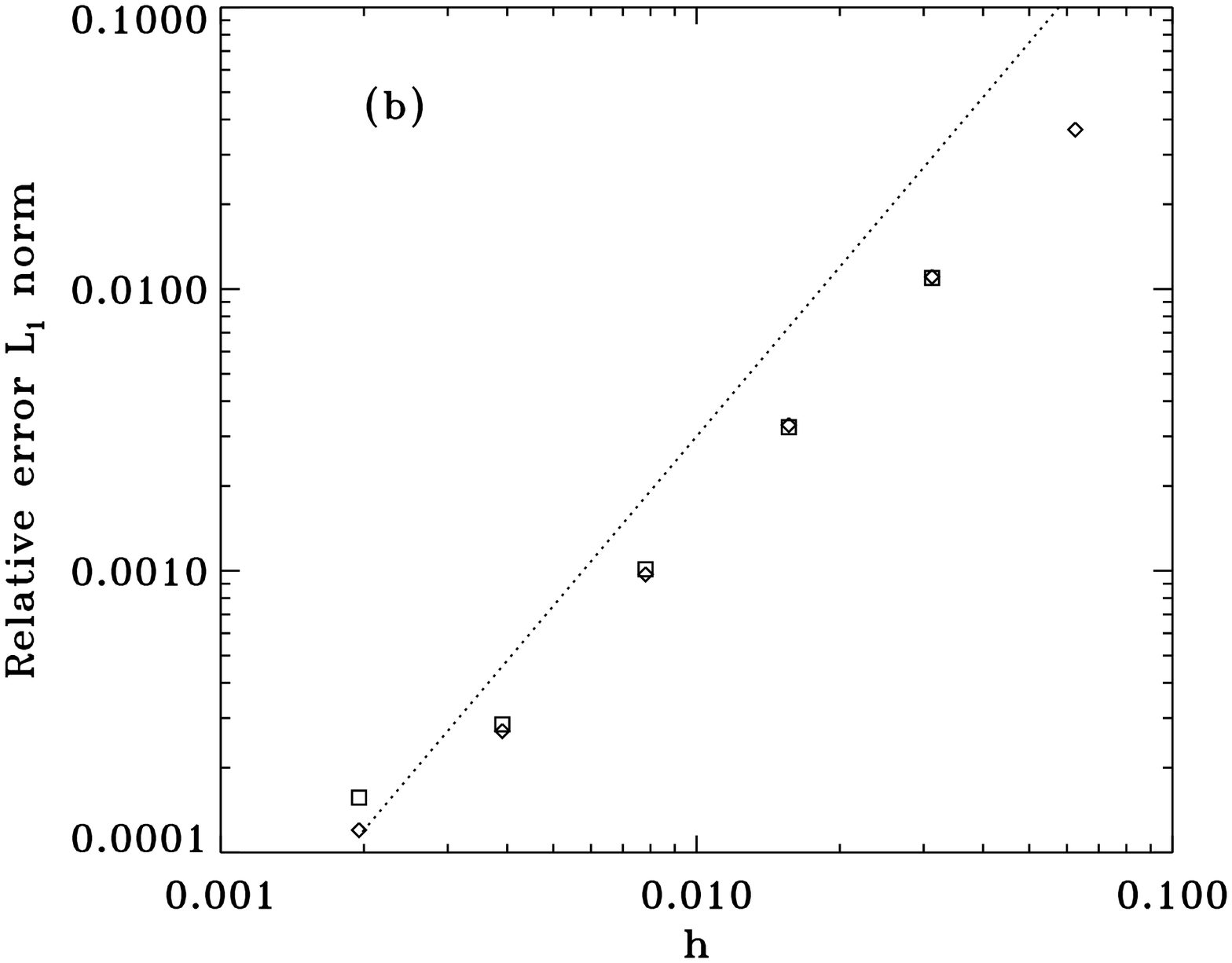}}
 \caption{Nonlinear diffusion of an initial step function with AMR, the refinement criterion based on radiative energy gradients. ({\it a}) Radiative energy profiles at time $t= 1.4\times 10^{-2}$ (square - numerical solution, "exact" solution in red, run with 2048 cells). The AMR levels (right axis) are plotted in blue. ({\it b})  L$_1$ norm of the error as a function of $h=1/\Delta x$, without AMR (diamond) and with AMR (squares), up to an effective resolution of 512 cells (the error is plotted as a function of the minimum mesh spacing, corresponding to the maximum resolution). The dotted line shows the evolution of the error as a function of $h^2$.}
	\label{diff_non_lin}
\end{figure}

\subsection{Matter-radiation coupling test}
Another conventional test is the matter-radiation coupling. Consider a static, uniform, absorbing fluid initially out of thermal balance, in which the radiation energy $E_\mathrm{r}$ dominates and is constant. An analytic solution can be obtained for the time evolution of the gas energy $e$, by solving the ordinary differential equation \citep{Turner_stone_01}
\begin{equation}
\frac{de}{dt}=c\sigma E_\mathrm{r}-4\pi \sigma B(e).
\end{equation}
We performed two tests, with two initial gas energies, $e=10^{10}$ erg cm$^{-3}$ and $e=10^{2}$ erg cm$^{-3}$.  In both tests, the following quantities are taken constant: the radiative energy $E_\mathrm{r} = 1\times 10^{12}$ erg cm$^{-3}$, the opacity $\sigma =4 \times 10^{-8}$ cm$^{-1}$, the density $\rho = 10^{-7}$ g cm$^{-3}$, the mean molecular weight $\mu=0.6$, and the adiabatic index $\gamma=5/3$. Figure \ref{coupling_test} shows the evolution in time of the gas energy for the analytic solution (red line) and the numerical solution (squares).  In the first calculations, where the initial gas temperature is greater than the radiative temperature, we used a variable time step $\Delta t $ that increases with time, starting from $10^{-20} $ s.  This good sampling gives very good results. In the second case, we used a constant time step $\Delta t = 10 ^{-12}$ s. Although the sampling is bad at early times and longer than the cooling time, numerical solutions always match the analytic one. This validates our linearization of the emission term ($a_\mathrm{R}T^4$) in equation (\ref{lin_tp}).

\begin{figure}[thb]
  \centering
  \resizebox{\hsize}{!}{ \includegraphics{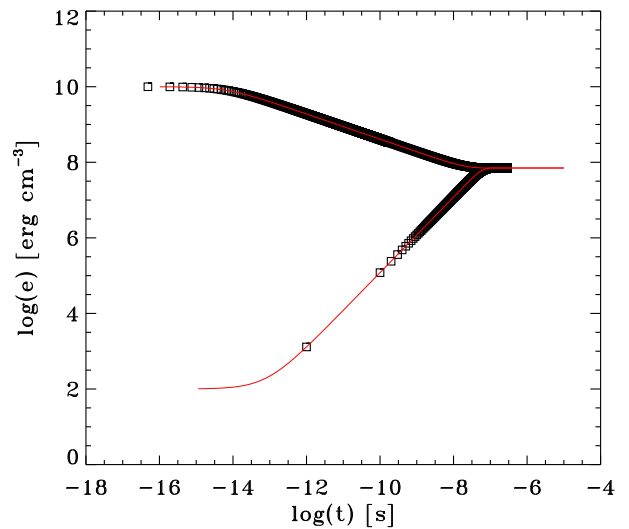}}
  \caption{Matter-radiation coupling test. The radiative energy is kept constant, $E_r = 1\times 10^{12}$ erg cm$^{-3}$, whereas the initial gas energies are out of thermal balance ($e=10^{2}$ erg cm$^{-3}$ and $e=10^{10}$ erg cm$^{-3}$). Numerical (square) and analytic (red curve) evolutions of the gas energy are given as a function of time.}
\label{coupling_test}
\end{figure}

\begin{figure*}[thb]
  \centering
  \includegraphics[width=8.75cm,height=7.25cm]{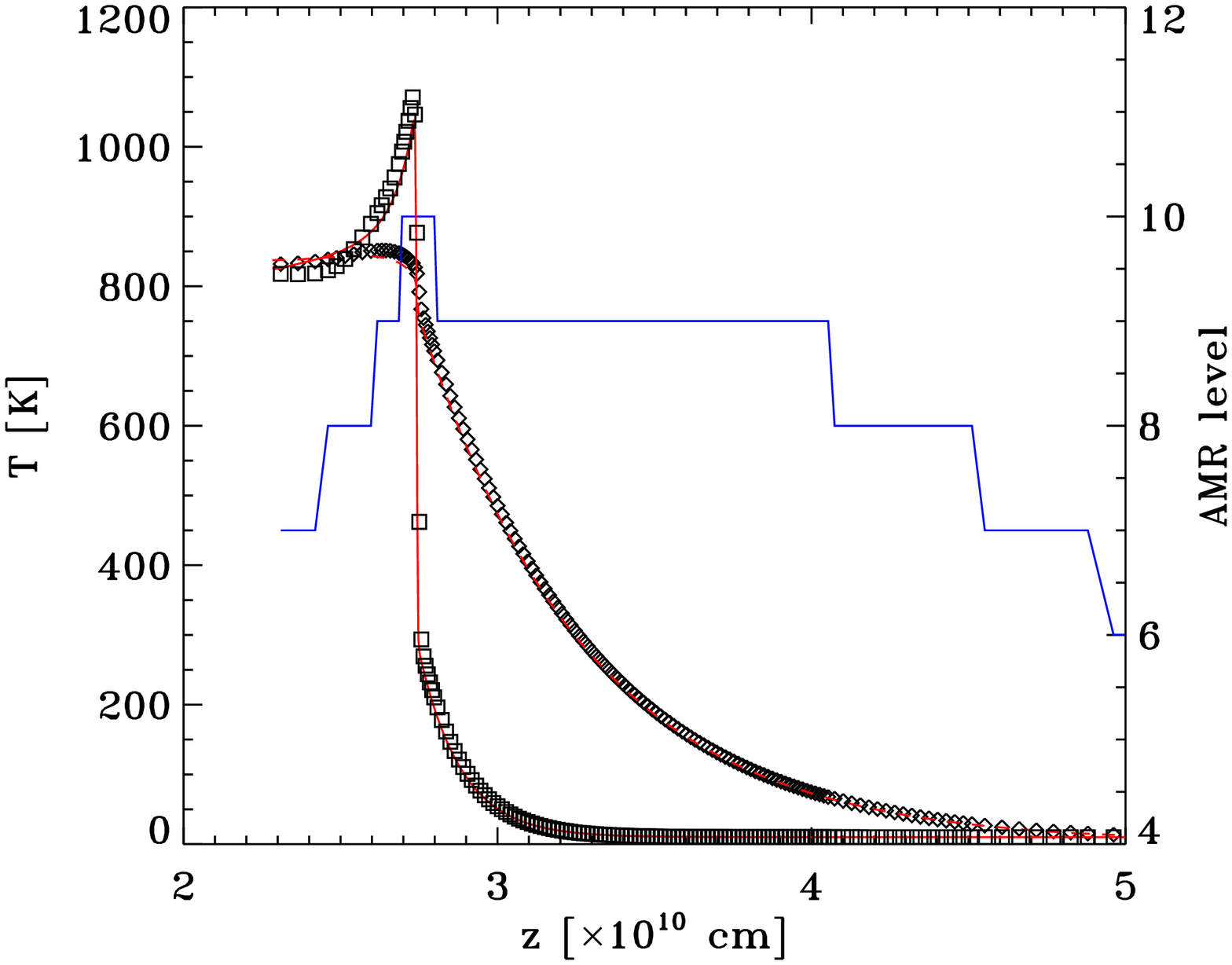}
  \hspace{10pt}
  \includegraphics[width=8.75cm,height=7.25cm]{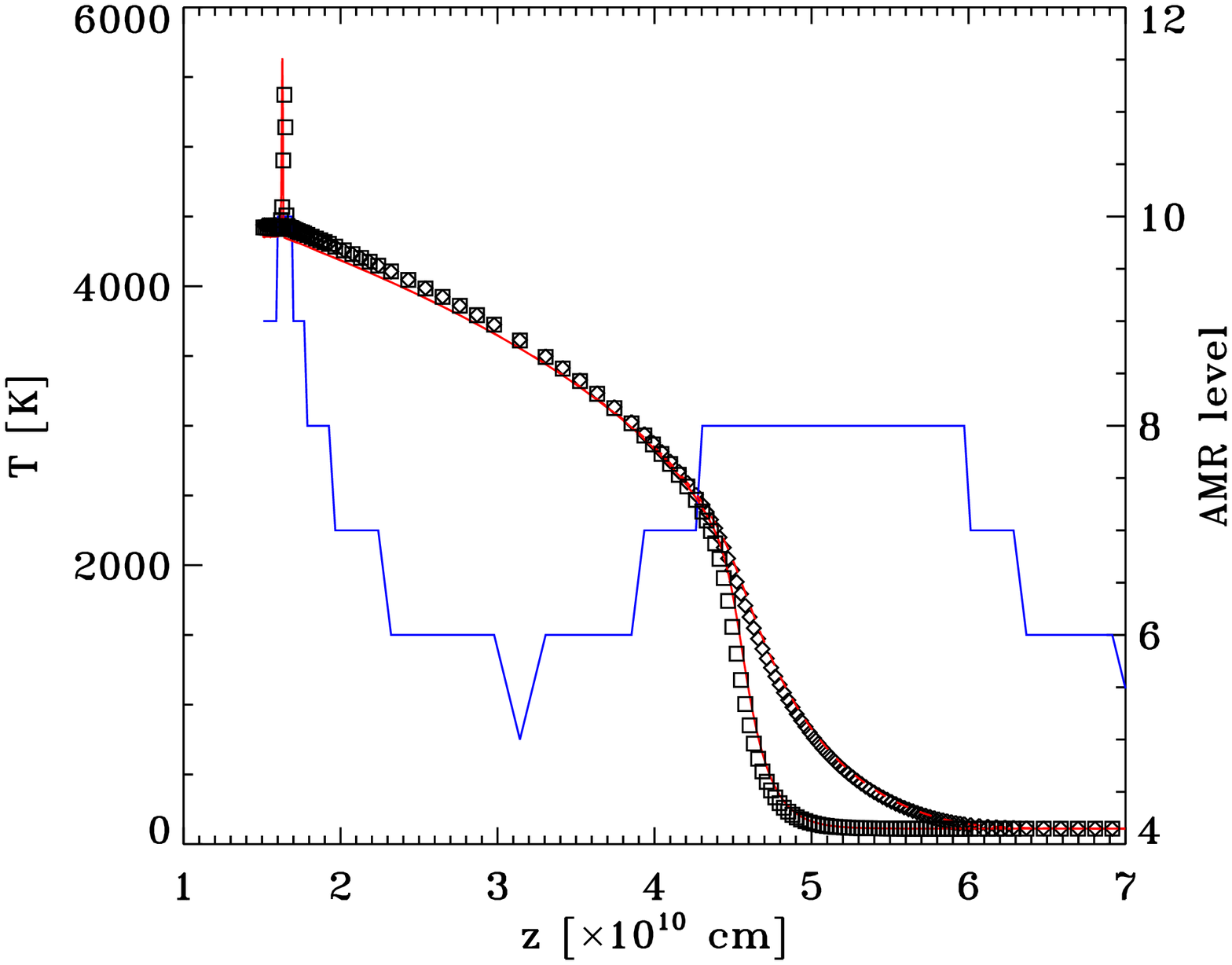}
  \caption{{\it Left}: Temperature profiles for a subcritical shock with piston velocity $v=6$ km s$^{-1}$, at time $t=3.8\times 10^{4}$ s.  {\it Right}: Temperature profiles for a supercritical shock with piston velocity $v=20$ km s$^{-1}$, at time $t=7.5\times 10^{3}$ s. In both cases, the temperatures are displayed as a function of $z=x-vt$. The squares represent the gas temperature and the diamonds the radiative temperature. The red curves represent the gas and radiative temperatures obtained with a calculation using 2048 cells, which we take as the "exact" solution. The AMR levels (blue line - right axis) are overplotted.}
\label{rshock_amr}
\end{figure*}
\subsection{1D full RHD tests: Radiative shocks}

Testing the numerical method for radiative shock calculations is a last important step that every code attempting to integrate RHD equations should perform \citep{Hayes_Norman_03,Whitehouse_Bate_2006,Gonzalez_2007}. Following the \cite{Ensman_94} initial conditions, we tested our routine for sub- and super-critical radiative shocks.

Initial conditions are as follows: uniform density  $\rho_0=7.78\times10^{-10}$ g cm$^{-3}$ and temperature T$_0$=10 K. The box length is L=$7\times 10^{-10}$cm, the opacity is constant ($\sigma =3.1\times10^{-10}$ cm$^{-1}$), $\mu=1$ and $\gamma=7/5$. The 1D homogeneous medium moves with a uniform speed (piston speed) from right to left and the left boundary is a wall. The shock is generated at this boundary and travels backwards. The piston velocity varies, producing sub- or super-critical radiative shocks. The AMR is used, and the refinement criterion is based on the density and radiative energy gradients (30\%), the grid has 32 coarse cells and we use five levels of refinement. We use the Minerbo flux limiter. The time step is given by  the hydrodynamics CFL for the explicit and implicit schemes.

Figure \ref{rshock_amr} shows the gas and radiative  temperatures for sub- and super-critical radiative shocks, as a function of $z=x-vt$, where $v$ is the piston's velocity. The AMR is used in both calculations. The squares represent the gas temperature and the diamonds the radiative temperature. The red curves represent the gas and radiative temperatures obtained with a calculation using 2048 cells, which we take as the "exact" solution. 
The subcritical shock is obtained with a piston velocity $v=6$ km s$^{-1}$, whereas the supercritical shock is obtained with $v=20$ km s$^{-1}$. In both tests, the occurrence of an extended, non-equilibrium radiative precursor is obvious. 
As expected, pre- and post-shock gas temperatures are equal in the supercritical case. 

For the subcritical case, the postshock gas temperature is given by \citep{Ensman_94,Mihalas_book}
\begin{equation}
T_2\approx \frac{2(\gamma-1)v^2}{\mathcal{R}(\gamma+1)^2},
\end{equation}
where $\mathcal{R}=k/\mu m_\mathrm{H}$ is the perfect gas constant. For our initial setup, this analytic estimate gives $T_2\sim 810$ K. Numerical calculations give $T_2\sim 825$ K at time $t=3.8\times 10^{4}$ s,  which agrees with the analytic estimate comparable to values obtained with more accurate methods \citep{Gonzalez_2007}. The characteristic temperature $T_{-}\sim 275$ K immediately in front of the shock agrees very well with the analytic estimate \citep{Mihalas_book}
\begin{equation}
T_{-}\approx \frac{\gamma-1}{\rho v R }\frac{2\sigma_\mathrm{R}T_2^4}{\sqrt{3}}\sim 276 \hspace{2pt}\mathrm{ K}.
\end{equation}
This means that in front of the shock, the gas internal energy flux flowing downstream is equal to the radiative flux flowing upstream. The entire radiative energy is absorbed upstream and contributes to heat the upstream gas. 
Similarly, the spike temperature $T_{+}\sim 1038$ K  also agrees well with the analytic estimate of \cite{Mihalas_book}
\begin{equation}
T_{+} \approx T_2+ \frac{3-\gamma}{\gamma+1}T_{-}\sim 980  \hspace{2pt}\mathrm{ K}.
\end{equation}

We note that the AMR scheme enables us to accurately describe the gas temperature spike at the shock. The medium around the spike is optically thin, and the numerical resolution in this region is therefore of crucial importance. The spike's amplitude varies according to the model used for radiation and to the effective numerical resolution. Thanks to the AMR scheme, the spike's amplitude is larger in the supercritical case, but not as large as those obtained with M1 or VTEF models \citep{Hayes_Norman_03,Gonzalez_2007}. However, this last test shows the ability of our time-splitting method to integrate the RHD equations.

\begin{table*}[t]      
\center
\begin{tabular}{ccccccccc}
\hline
\hline
Reference &$R_\mathrm{fc}$ & $M_\mathrm{fc}$ &$\dot{M}$             & $L_\mathrm{acc}$ & $T_\mathrm{c}$ & $T_\mathrm{fc}$ & $S_\mathrm{c}$  &$\alpha_\mathrm{acc}$\\
 &                      (AU) &   (M$_\odot$)       & (M$_{\odot}/$yr) &   (L$_\odot$)         &  (K) & (K) & (erg K$^{-1}$ g$^{-1}$) & \\
\hline
This work &\small
  8  &  $2.1 \times 10^{-2}$  &  $3.7 \times 10^{-5}$  & 0.014  &  396  &  81  & $2.11\times 10^9$ &24\\ 
\cite{Masunaga_Miyama_Inutsuka_I_1998ApJ}&\small
  $\sim 8$  &  $ \sim 10^{-2}$  &  $ \sim10^{-5}$  & 0.002  & $\sim 200$  &  60  & $2.08\times 10^9$ & 6\\ 
\cite{Commercon_2010} &\small
  7  &  $2.31 \times 10^{-2}$  &  $3 \times 10^{-5}$  & 0.015  &  419  &  70  & $2.02\times 10^9$ &19\\ 
\cite{Larson_1969} &\small
  $\sim$4  &  $\sim 1 \times 10^{-2}$  &  - & - &  170 &  -  & $ - $ &- \\ 
   \hline
\end{tabular} 
\caption{Summary of first core properties at time $t=$1.012 $t_\mathrm{ff}$ and  $\rho_c=2.7\times 10 ^{-11}$ g cm$^{-3}$. $R_\mathrm{fc}$, $M_\mathrm{fc}$ and $T_\mathrm{fc}$ give the radius and mass of the first core, and the temperature at the first core border respectively. The mass accretion rate $\dot{M}$ and accretion luminosity $L_\mathrm{acc}=GM_\mathrm{fc}\dot{M}/R_\mathrm{fc}$ are also computed at the first core border. $T_\mathrm{c}$ and $S_\mathrm{c}$ give the central temperature and entropy. $\alpha_\mathrm{acc}$ is a typical accretion parameter.
The comparative values  are roughly estimated at $\rho_\mathrm{c}\sim\times 10 ^{-11}$ g cm$^{-3}$ in \cite{Masunaga_Miyama_Inutsuka_I_1998ApJ}, at $\rho_\mathrm{c}= 10 ^{-10}$ g cm$^{-3}$ in \cite{Commercon_2010} and at $\rho_\mathrm{c}= 2\times10 ^{-10}$ g cm$^{-3}$ in \cite{Larson_1969}.}
\label{a050_norot}
\end{table*}
 
 \begin{figure*}[htb]
  \centering
  \includegraphics{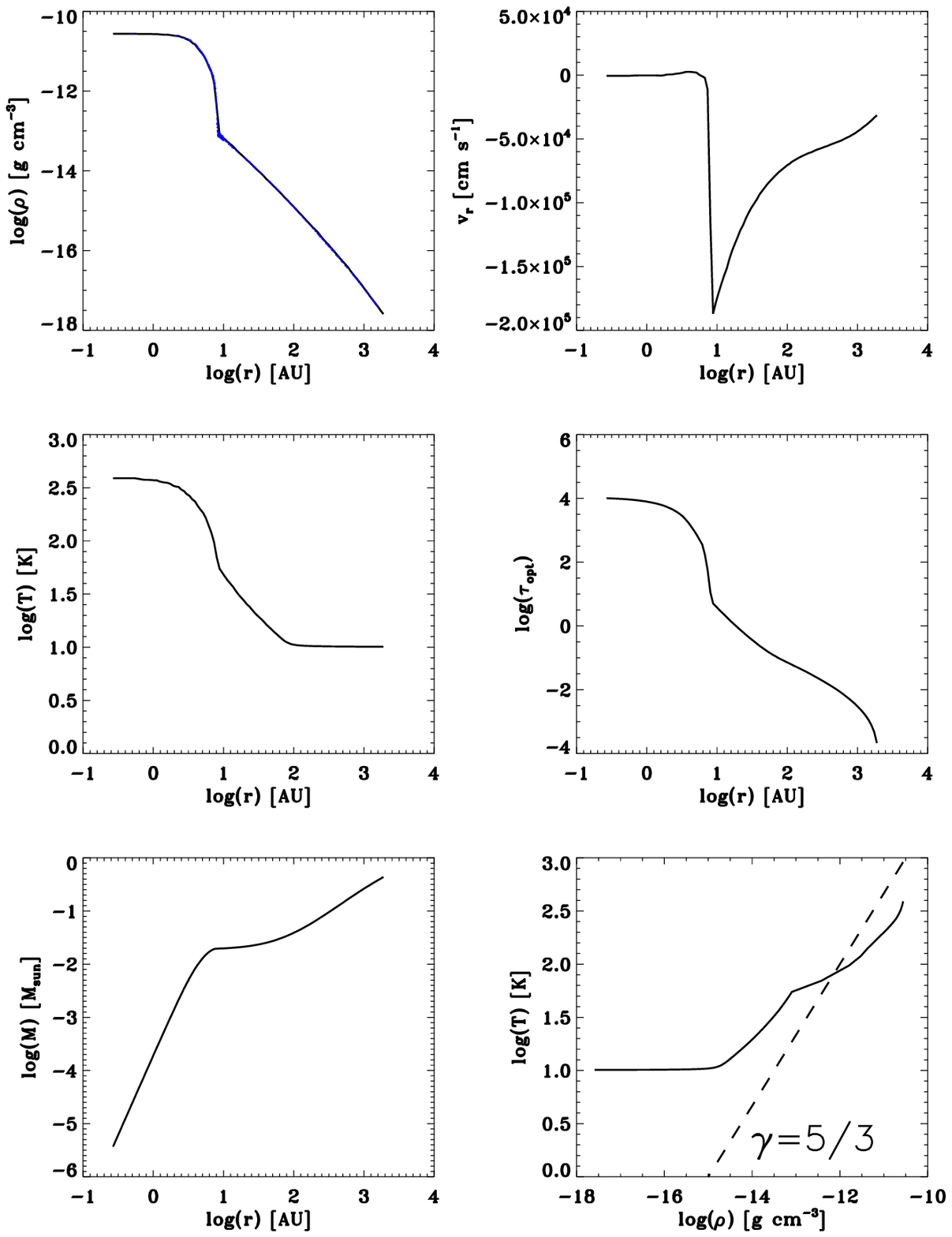}
\caption{Profiles of density,  radial velocity, temperature, optical depth, and integrated mass as a function of the radius and the temperature as a function of  density in the 3D computational domain. All values are computed at time $t=$1.012 $t_\mathrm{ff}$.}
\label{profile_a050_norot}
\end{figure*}

\subsection{3D dense-core-collapse calculations without rotation}

In this section, we perform calculations of a 1 M$_\odot$ dense-core collapse without rotation, using our grey FLD solver. We compare our FLD results   for a model without initial rotation with the  ones obtained by \cite{Masunaga_Miyama_Inutsuka_I_1998ApJ} and with our results obtained with a 1D code \citep[see][]{Commercon_2010}. We also qualitatively compare our results with the pioneered ones of \cite{Larson_1969} and \cite{Winkler_Newman_1980}. This latter test provides a validation in 3D for star-formation purposes.


\subsubsection{Initial conditions}

To facilitate the comparison with other studies, we used the same initial conditions as in \cite{Commercon_2008} and in Paper II and the Lax Friedrich Riemann solver. We chose highly gravitationally unstable  initial conditions. The initial sphere is isothermal, $T_0=10$ K, and has a uniform density $\rho_0=1.38\times 10^{-18}$ g cm$^{-3}$.  The ratio $\alpha$ between initial thermal energy and gravitational energy is  $\alpha\sim 0.50$. The initial radius is $R_0=7.07\times 10^{16}$ cm. The theoretical free-fall time is $t_\mathrm{ff}=57$ kyr. The initial isothermal sound speed is $c_{s0}\sim 0.19$ km s$^{-1}$ and $\gamma=5/3$. 
The outer region of the sphere
is at the same temperature as the core  temperature, but is  100 times less  dense.  The sphere  radius is equal to a quarter of
the box length to minimize border effects.
 
 We use   the set  of
opacities given  by \cite{Semenov_et_al_2003A&A} for  low temperature
($<1000$ K), which 
we compute as a function of the gas temperature and density.  For each
cell  we perform  a  bilinear interpolation  on  the mixed  opacities
table. Below 1500 K the opacities are dominated by grain (silicate,
iron, troilite, etc...).  \cite{Semenov_et_al_2003A&A} take into account the dependence of the evaporation temperatures of ice, silicates, and iron on the gas density. We here use spherical composite aggregate particles for the grain structure and topology and a normal iron content in the silicates, Fe/(Fe + Mg)=0.3.

 \subsubsection{Results}
   To resolve the Jeans length, we use $N_\mathrm{J}=10$ (i.e. 10 points per Jeans length) . \cite{Masunaga_Miyama_Inutsuka_I_1998ApJ} showed that  the first core properties are independent of the initial conditions for low-mass star formation. We can then compare our results with those obtained by  \cite{Masunaga_Miyama_Inutsuka_I_1998ApJ}, even if we use different initial conditions. We also compare our results with those obtained using a 1D spherical code \citep{Audit_2002} in \cite{Commercon_2010}.

Table \ref{a050_norot} summarizes the first core properties obtained at time $t=$1.012 $t_\mathrm{ff}$ with {\ttfamily RAMSES}. The first core radius and mass are qualitatively similar to the results obtained in other 1D Lagrangean calculations \citep[see][]{Masunaga_Miyama_Inutsuka_I_1998ApJ,Commercon_2010}, even though we use a completely different hydrodynamical scheme (e.g. no artificial viscosity, Eulerian, etc...). The first core radius is however a factor 2 greater than the one found in \cite{Larson_1969} and \cite{Winkler_Newman_1980}, who used simplified dust opacity models. Since the first core is mainly set by the opacity, this explains the differences.
  We define the first core radius as the radius at which the infall velocity is maximal. The accretion rate on the first core is typical of low-mass star formation, $\sim 10^{-5}$ M$_{\odot}/$yr. We note that the value $\alpha_\mathrm{acc}\sim 24$ is relatively high, with $\alpha_\mathrm{acc}$ defined as $\dot{M}=\alpha_\mathrm{acc}c^3_{s0}/G $ (where $c_{s0}$ the isothermal sound speed). This indicates that our collapse model is closer to the dynamical Larson-Penston collapse solution \citep{Larson_1969,Penston_1969} than to the classical SIS model of \cite{Shu_1977}, for which $\alpha_\mathrm{acc}\sim 0.975$. 

In Fig. \ref{profile_a050_norot} we show the profiles of density, radial velocity, temperature, optical depth, and integrated mass as a function of the radius and the temperature as a function of the density in the computational domain at time $t=$1.012 $t_\mathrm{ff}$. All quantities are mean values in the equatorial plane. In the density profiles, all cells of the calculations have been displayed (blue points). The spread in the density distribution is very small. The spherical symmetry is thus well conserved in the 3D calculations with {\ttfamily RAMSES}. We compare these profiles with those obtained in \cite{Commercon_2010}. The density jump between the first core border and the centre is of the same order of magnitude as for the 1D spherical case. The infall velocity at the shock is also comparable ($\sim 2$ km s$^{-1}$). 
The accretion shock takes place around $\tau\sim 5-10$, in the optically thick region. We do not see a jump in temperature through the accretion shock, which is a supercritical radiative shock. Eventually, we see from the temperature versus density plot that the thermal behaviour of the gas is not perfectly adiabatic in the central core. The first core is not fully adiabatic and is able to decrease its entropy level by radiating in the upstream material.
The slight kink in the curve at $T\sim 80$ K (log$(T)\sim1.7$) corresponds to ice evaporation in the opacity table. The opacity decreases abruptly, this is the reason why the cooling is more efficient in that region.

\section{Summary and perspectives}

We have developed a full radiation-hydrodynamics solver using the flux-limited diffusion approximation, which is integrated in the AMR {\ttfamily RAMSES}  code.  Our solver uses a time-splitting integrator scheme and combines explicit and implicit methods.
Each step of the integration is detailed in this work. The method was successfully tested in several conventional tests in 1D and 2D. We demonstrated that our method is second-order accurate in space, even when AMR is used. 
We also performed collapse calculations of a non-rotating dense core and successfully  compared our results with those  obtained by \cite{Masunaga_Miyama_Inutsuka_I_1998ApJ} and \cite{Commercon_2010}, which are based on different methods in 1D spherical codes. Our method has thus  been demonstrated to be robust and well suited for star formation. In Paper II we present detailed RHD calculations with a very high resolution of dense-core collapse in rotation. We showed that our method enables us to accurately handle the heating and cooling processes. 
Last but not least, we extended our method  to the radiation-magnetohydrodynamics flows in \cite{Commercon_2010L}. 

The next step following this work will be to tune our solver for adaptive time-stepping to make use of all benefits of the AMR in {\ttfamily RAMSES}. For example, the next stages of the collapse, the second collapse and the second core formation, require a huge amount of  numerical resolution and the dynamical timescale becomes much shorter. An adaptive time-step scheme is then suitable.  Another improvement is to use a multi-group approach in the radiation solver. Some attempts have been presented in the literature \citep[e.g.][]{Shestakov_2008}, but the computational cost remains too high nowadays compared to the grey model.

\begin{acknowledgements}
The calculations were performed at CEA on the DAPHPC cluster. The research of B.C. is supported by the postdoctoral fellowships from Max-Planck-Institut f\"{u}r Astronomie . The research leading to these results has received funding from the 
European Research Council under the European Community's Seventh 
Framework Programme (FP7/2007-2013 Grant Agreement no. 247060). BC also thanks Neal Turner for useful discussions. 
\end{acknowledgements}
\bibliographystyle{aa}
\bibliography{biblio1}
\begin{appendix}

\section{The super-time stepping versus the conjugate gradient\label{STS}}
In this appendix we present the super-time stepping (STS)  method. It is used to solve parabolic equation systems, like the conjugate gradient (CG) we used previously.  We implement the STS scheme into {\ttfamily RAMSES}. We compare the CG and the STS methods for the particular case of the 1D linear diffusion test presented in \S \ref{1D_diff_lin}.

\begin{figure*}[t]
  \centering
  \includegraphics[width=7.2cm,height=5.25cm]{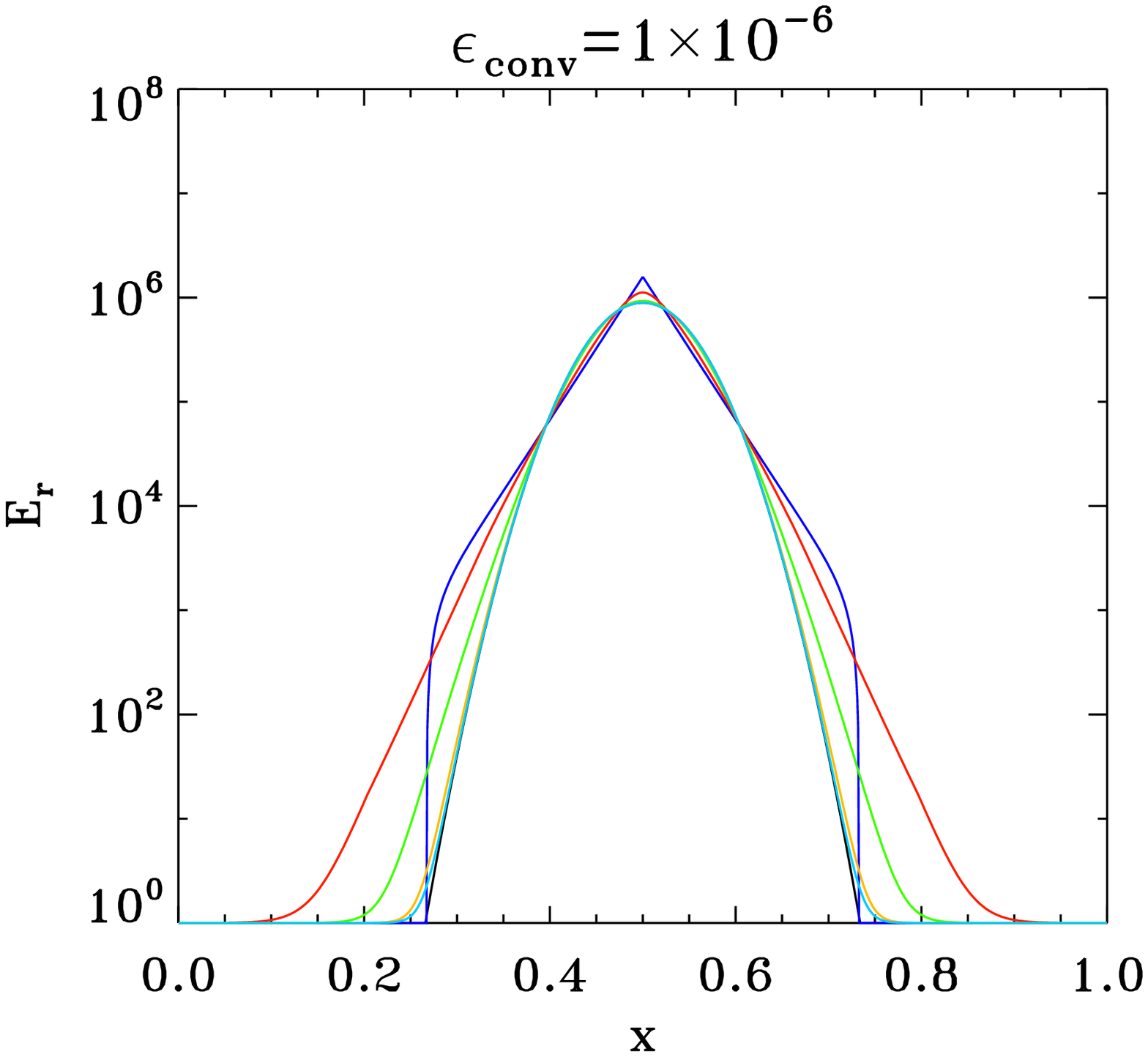}
  \includegraphics[width=7.2cm,height=5.25cm]{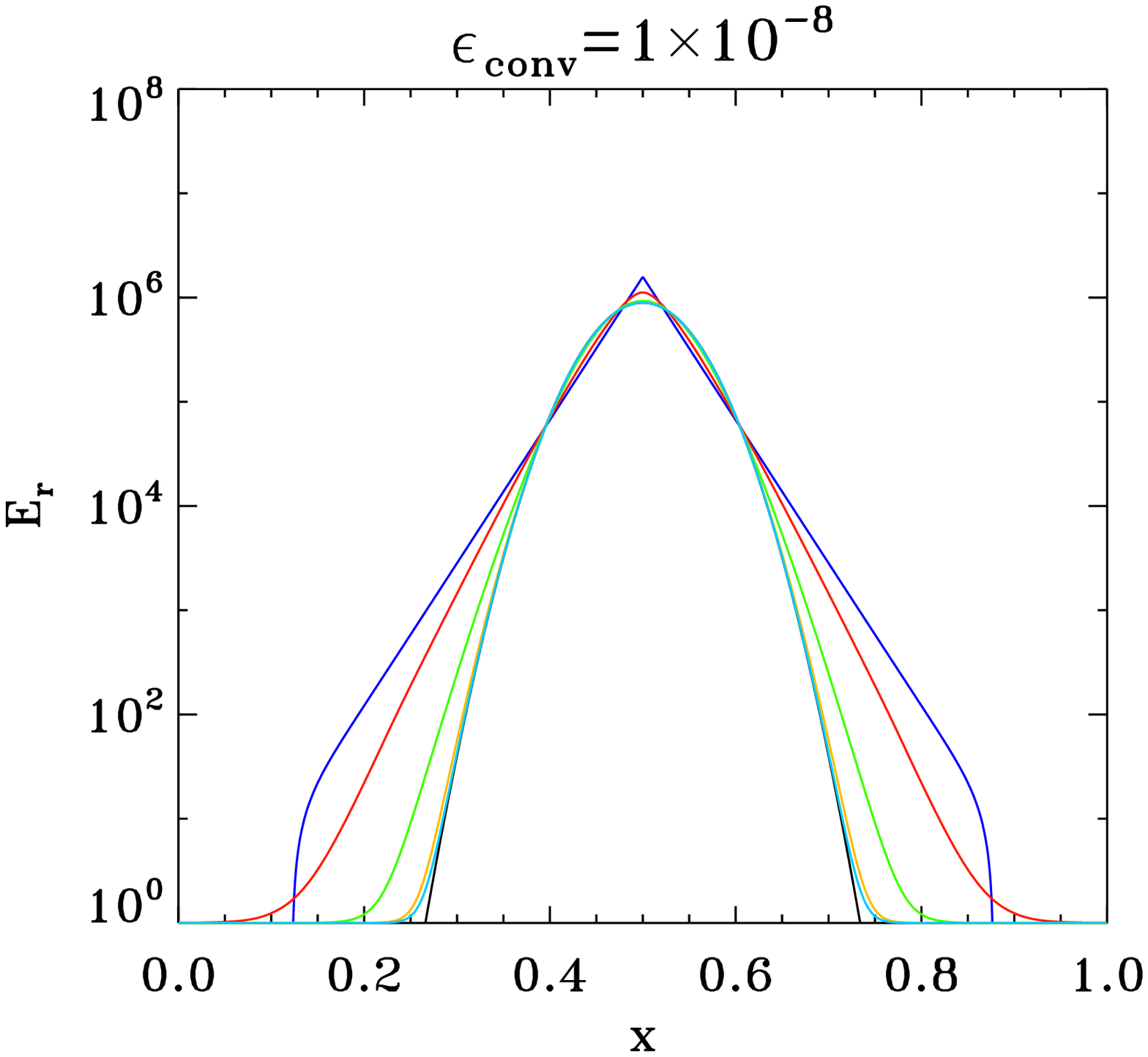}
  \includegraphics[width=7.2cm,height=5.25cm]{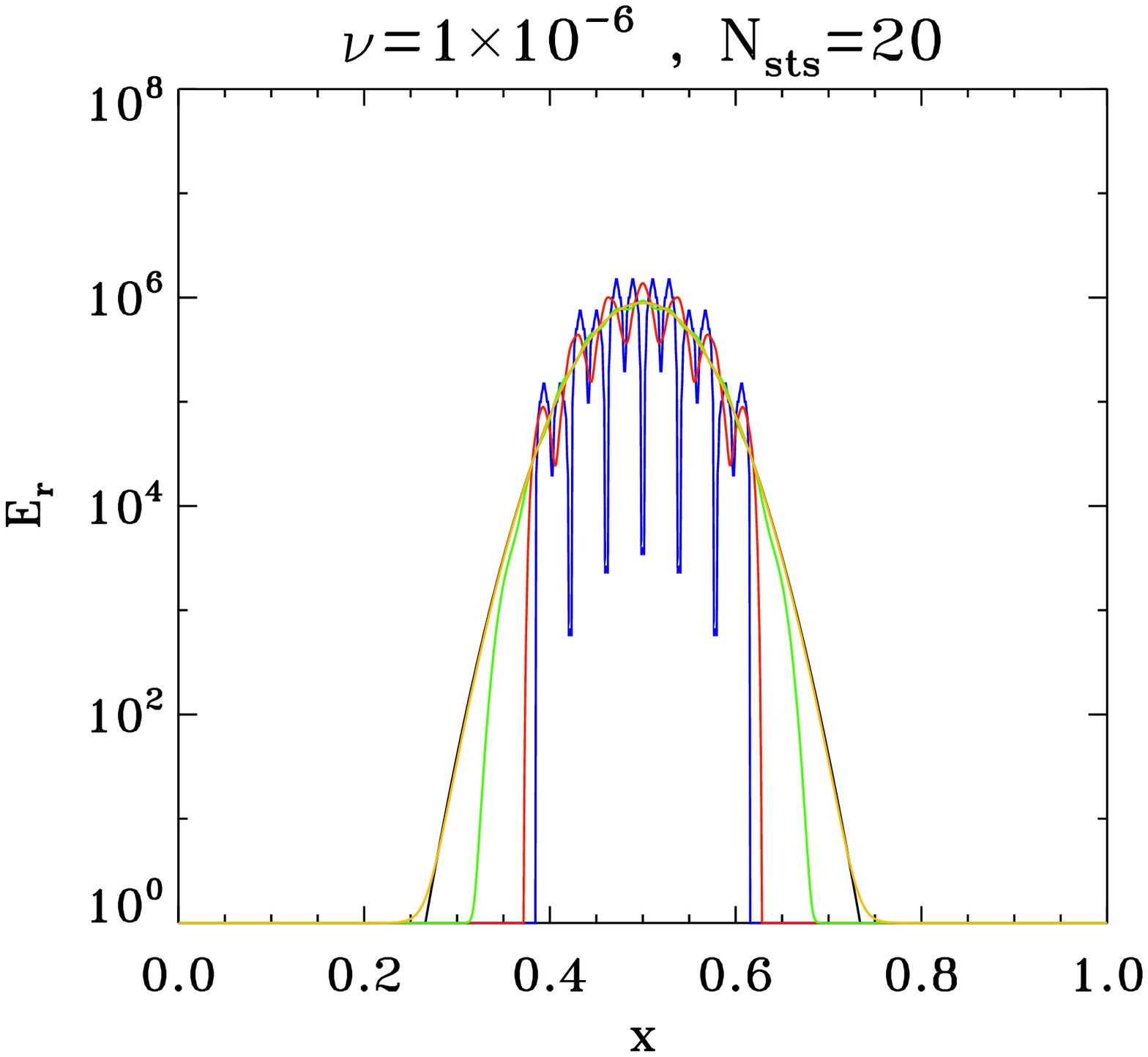}
  \includegraphics[width=7.2cm,height=5.25cm]{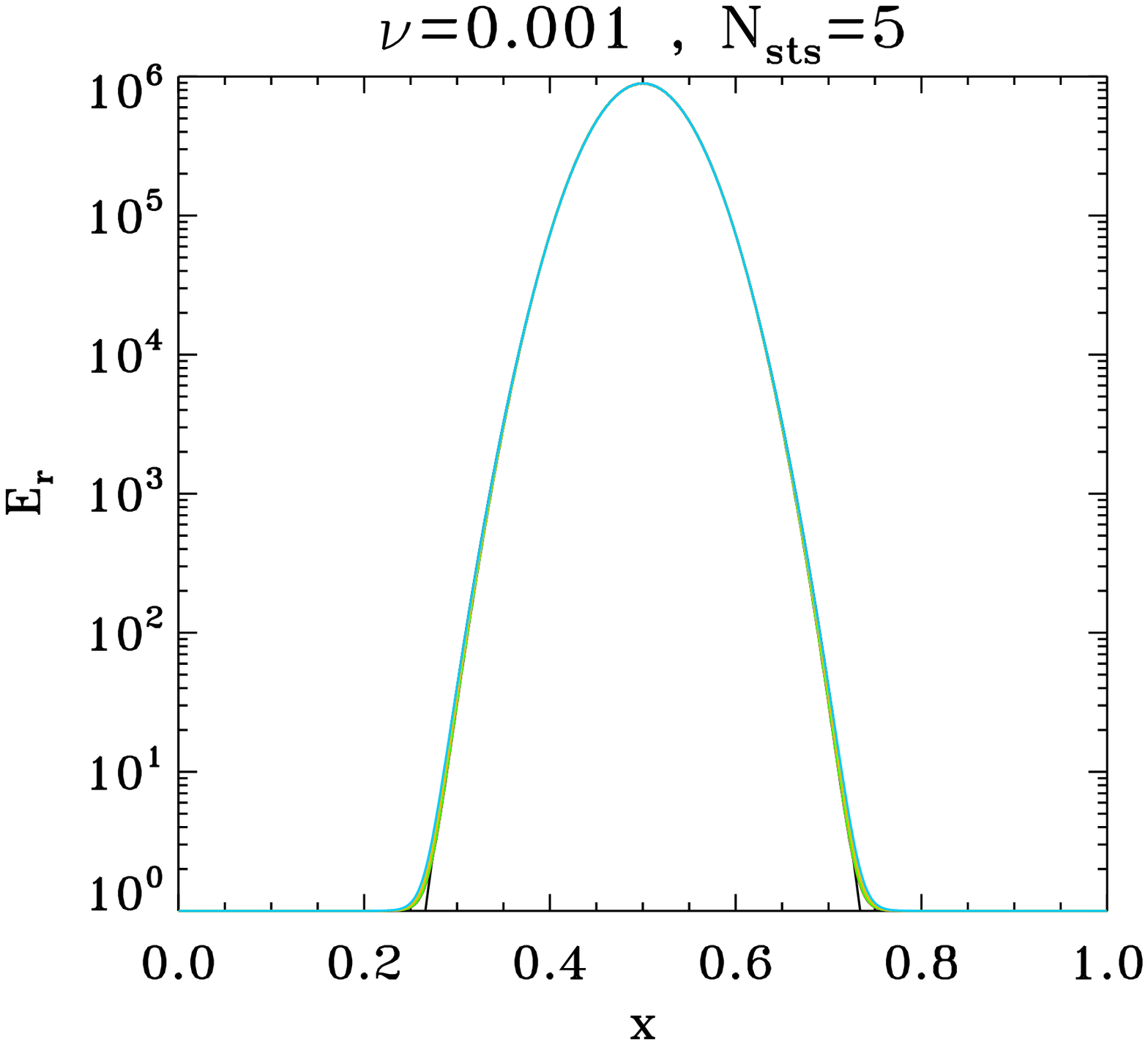}
  \includegraphics[width=7.2cm,height=5.25cm]{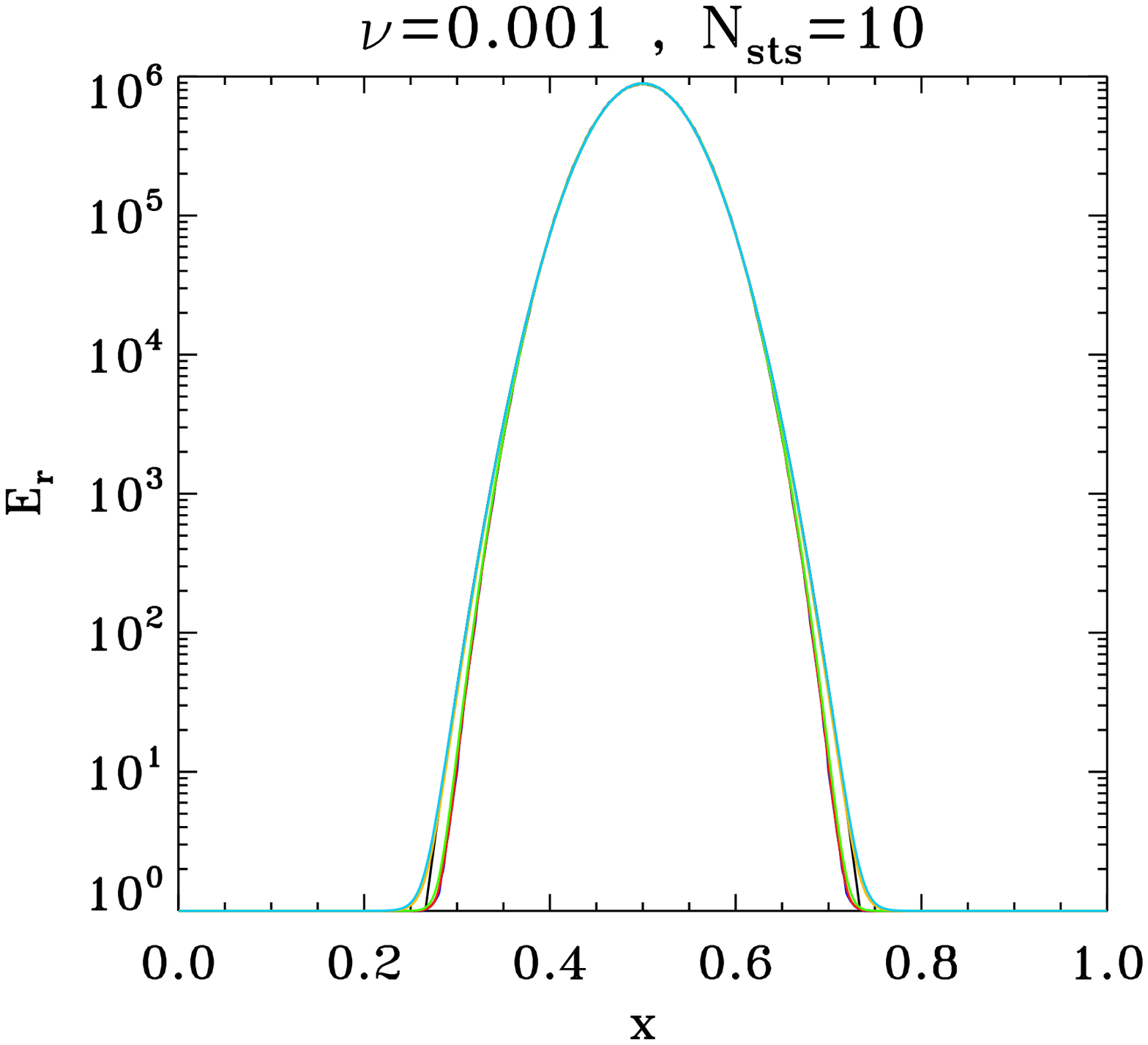}
  \includegraphics[width=7.2cm,height=5.25cm]{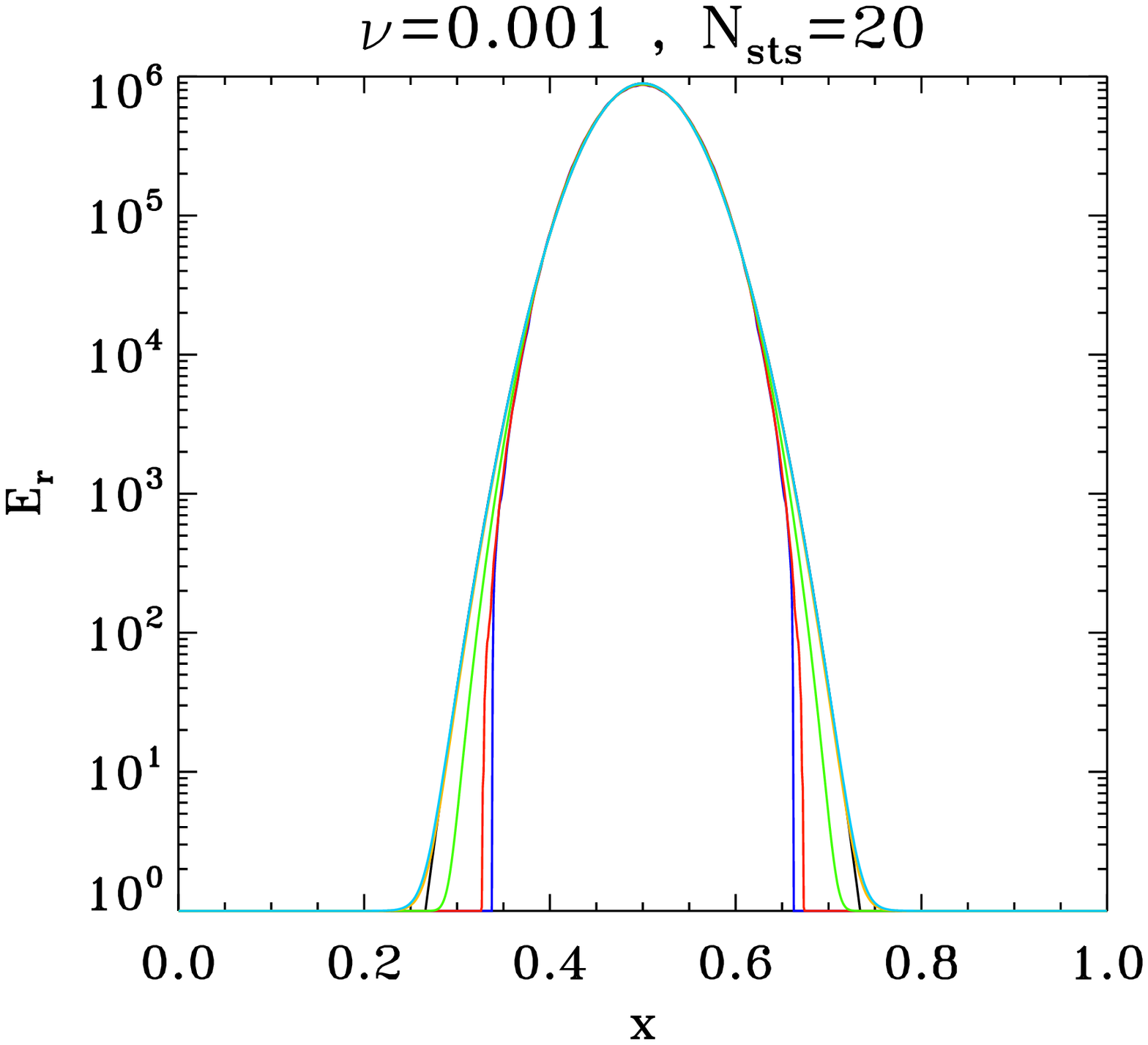}
  \includegraphics[width=7.2cm,height=5.25cm]{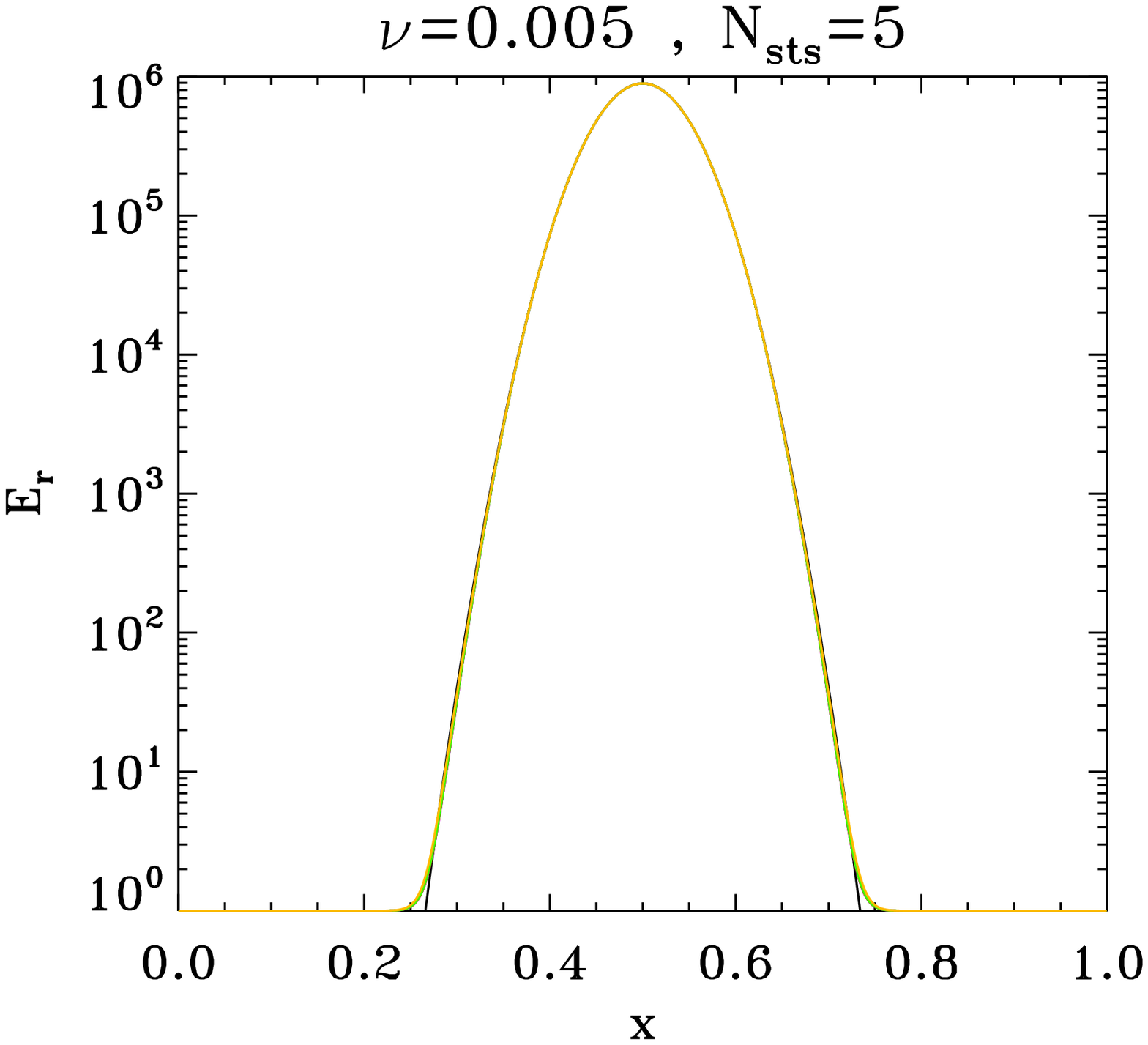}
  \includegraphics[width=7.2cm,height=5.25cm]{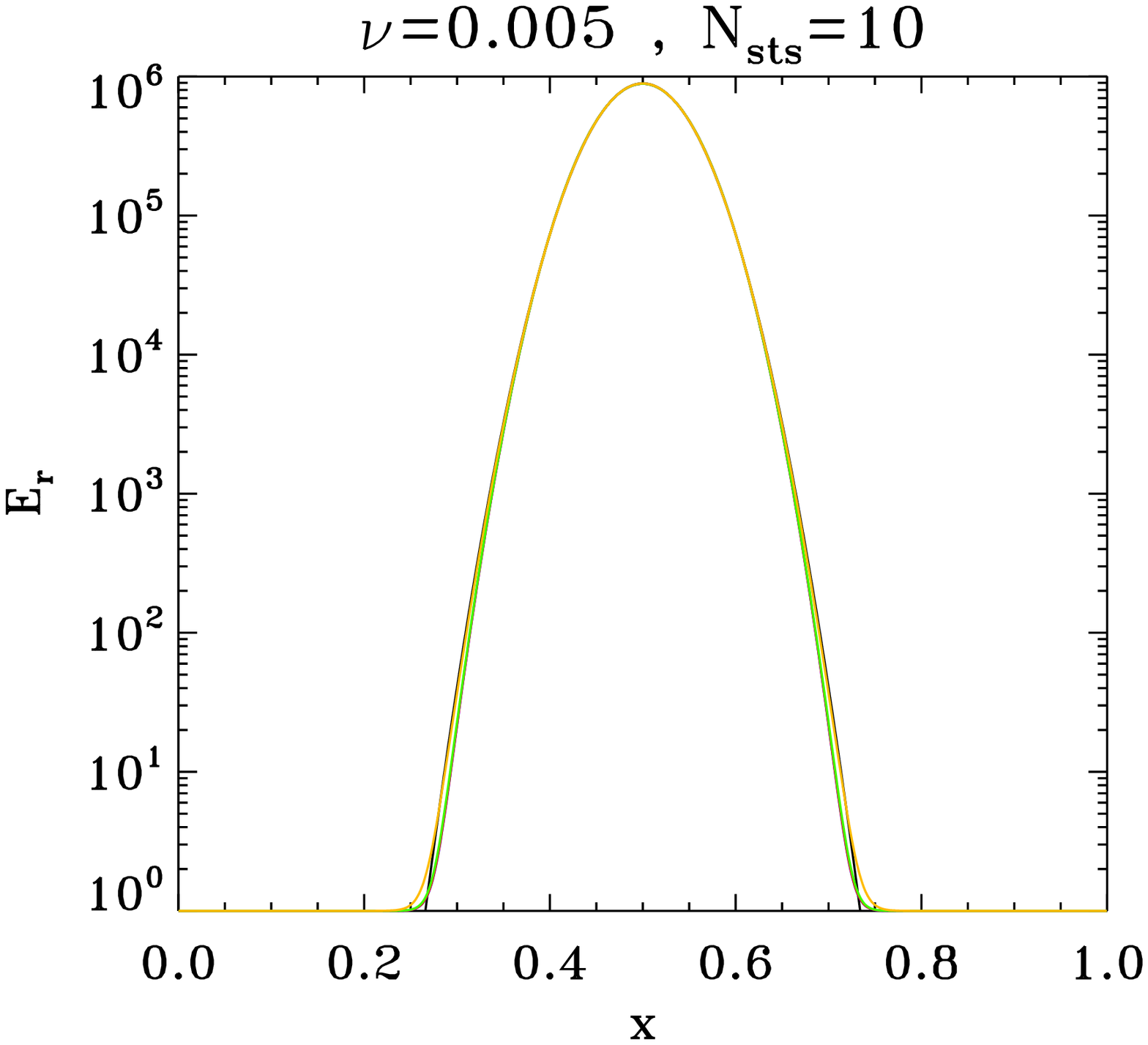}
  \caption{Comparison  of the numerical solutions using STS or CG with the analytic one (black line) at time $t=1\times 10^{-13}$ s. The color curves depict numerical solutions obtained with timestep $\Delta t$ equals to $1\times10^{-13}$ (blue), $5\times10^{-14}$ (red), $1\times10^{-14}$ (green), $1\times10^{-15}$ (yellow) and $1\times10^{-16}$ (cyan).}
\label{sts_cg_fig}
\end{figure*}

\subsection{The super-time stepping}
The STS is a very simple and effective way to speed up explicit time-stepping schemes for parabolic problems. The method has been recently rediscovered in \cite{Alexiades_1996}, but it remains relatively unknown in computational astrophysics \citep{Mignone_2007,Osullivan_2006}. The STS frees the explicit scheme from stability restrictions on the time-step. It can be very powerful in some cases and is easy to implement compared to implicit methods that involve matrix inversions. \\
\\
The STS is designed for time dependent problem such as
\begin{equation}
\frac{dU}{dt}(t)+ AU(t)=0,
\label{sts_eq}
\end{equation}
where $A$ is a square, symmetric positive definite matrix. Equation \ref{sts_eq} is rewritten with the corresponding standard explicit scheme
\begin{equation}
U^{n+1}=U^n-\Delta tAU^n,
\end{equation}
The explicit scheme is subject to the restrictive stability condition
\begin{equation}
\rho(\mathbb{I} -\Delta t A)<1,
\end{equation}
where $\rho(\cdot)$ denotes the spectral radius. The equivalent CFL condition is
\begin{equation}
\Delta t< \Delta t_\mathrm{expl}=\frac{2}{\lambda_\mathrm{max}},
\end{equation}
where $\lambda_\mathrm{max}$ stands for the highest eigenvalue of $A$. For the 1D heat equation $\partial u/\partial t=\chi \Delta u$, discretized by standard second-order differences on a uniform mesh, we have $\lambda_\mathrm{max}=4\chi \Delta x^2$ ($\Delta t_\mathrm{expl}=\Delta x^2/2\chi$).

In the STS method, the restrictive stability condition is relaxed by requiring the stability at the end of a cycle of $N_\mathrm{sts}$ time-steps instead of requiring stability at the end of each time step $\Delta t$. It leads to a Runge-Kutta-like method with $N_\mathrm{sts}$ stages. Following \cite{Alexiades_1996}, we introduce a {\it superstep} $\Delta T=\sum_{j=1}^{N_\mathrm{sts}} \tau_j$ consisting of $N_\mathrm{sts}$ {\it substeps} $\tau_1,\tau_2,\cdot\cdot\cdot,\tau_{N_\mathrm{sts}}$. The idea is to ensure stability over the superstep $\Delta T$, while trying to maximize its duration. The inner values, estimated after each $\tau_j$, should only be considered as intermediate calculations. Only the values at the end of the superstep approximate the solution of the problem.\\
\\
The new algorithm can be written as
\begin{equation}
U^{n+1}=\left(\prod_{j=1}^{N_\mathrm{sts}}(\mathbb{I}-\tau_jA)\right)U^n,
\end{equation}
and the corresponding stability condition is 
\begin{equation}
\rho\left(\prod_{j=1}^{N_\mathrm{sts}}(\mathbb{I}-\tau_jA)\right)<1.
\end{equation}
In order to find $\Delta T$ as high as possible, the properties of Chebyshev polynomials are exploited,  providing a set of optimal values for the substeps given by
\begin{equation}
\tau_j=\Delta t_\mathrm{expl}\left[(-1+\nu_\mathrm{sts})\mathrm{cos}\left(\frac{2j-1}{N_\mathrm{sts}}\frac{\pi}{2} \right) +1+\nu_\mathrm{sts}\right]^{-1},
\end{equation}
where $\nu_\mathrm{sts}$ is a damping factor that should satisfy $0<\nu_\mathrm{sts}<\lambda_\mathrm{min}/\lambda_\mathrm{max}$. The superstep $\Delta T$ is given by
\begin{equation}
\Delta T=\sum_{j=1}^{N_\mathrm{sts}} \tau_j=\Delta t_\mathrm{expl}\frac{N_\mathrm{sts}}{2\nu_\mathrm{sts}^{1/2}}\left[\frac{(1+\nu_\mathrm{sts}^{1/2})^{2N_\mathrm{sts}}-(1-\nu_\mathrm{sts}^{1/2})^{2N_\mathrm{sts}}}{(1+\nu_\mathrm{sts}^{1/2})^{2N}+(1-\nu_\mathrm{sts}^{1/2})^{2N_\mathrm{sts}}}\right].
\end{equation}
Note that $\Delta T\rightarrow N_\mathrm{sts}^2\Delta t_\mathrm{expl}$ as $\nu_\mathrm{sts}\rightarrow 0$. The method is unstable in the limit $\nu_\mathrm{sts}=0$. The STS method is thus almost $N_\mathrm{sts}$ times faster than the standard explicit scheme. When $\Delta T$ is taken to be the advective (CFL) time step  $\Delta t$ while coupling with the hydrodynamics, the STS requires only approximately $(\Delta t/\Delta t_\mathrm{expl})^{1/2}$ iterations rather than $\Delta t/\Delta t_\mathrm{expl}$ with an explicit scheme. 

\subsection{The STS implementation for the FLD equation}

The STS scheme replaces the implicit radiative scheme presented in Sect. \ref{imp_part}. Equations of system (\ref{imp_schema}) written with an explicit scheme become

\begin{equation}
\left\{\begin{aligned}
\frac{C_v T^{n+1} - C_v T^{n}}{\Delta t} & = & - \kappa_\mathrm{P}^n\rho^n \mathrm{c}(a_\mathrm{R}(T^{n})^4 - E_\mathrm{r}^{n}) \\ 
\frac{E_\mathrm{r}^{n+1} - E_\mathrm{r}^{n}}{\Delta t} & = &   \nabla \frac{\mathrm{c}\lambda^n}{\kappa_\mathrm{R}^n\rho^n}\nabla E_\mathrm{r}^{n} + \kappa_\mathrm{P}^n\rho^n \mathrm{c}(a_\mathrm{R}(T^{n})^4 - E_\mathrm{r}^{n})
\end{aligned}
\right.  ,
\label{exp_schema}
\end{equation}

The explicit time step $\Delta t_\mathrm{expl}$ is estimated using values at time $n$. The next step consists of estimating values of $N_\mathrm{sts}$ and $\nu_\mathrm{sts}$, the latter depending on the spectral properties of $A$. However, as mentioned in \cite{Alexiades_1996}, it is not required to have a precise knowledge of the spectral properties for the method to be robust. $N_\mathrm{sts}$ and $\nu_\mathrm{sts}$ are thus arbitrary chosen by the user. Instead of executing one time step of length $\Delta t_\mathrm{expl}$, one executes supersteps of length $\Delta T$. $N_\mathrm{sts}$ substeps $\tau_1,\tau_2,\cdot\cdot\cdot,\tau_{N\mathrm{sts}}$ are thus performed  without outputing until the end of each superstep. When the STS is coupled to the hydrodynamics solver, the cycle is repeated until the time step, given by the hydrodynamical CFL condition, is reached. Superstep $\Delta T$ and substeps $\tau_i$ are re-estimated at the end of each cycle. 

\subsection{Comparison with the conjugate gradient method}
To compare the STS with the CG algorithm we used throughout, we consider the test case presented in Sect. \ref{1D_diff_lin}. The equation to integrate is simply
\begin{equation}
\frac{\partial E_\mathrm{r}}{\partial t} - \nabla \cdot\left(\frac{c}{3 \rho \kappa_\mathrm{R}} \nabla E_\mathrm{r}\right) = 0.
\end{equation}
The initial setup is identical to those in Sect. \ref{1D_diff_lin}. It consists of an initial pulse of radiative energy in the middle of the box. 
We present here calculations made with either the STS method or the CG algorithm. In both cases, CG and STS are applied over an arbitrary time step $\Delta t$ that simulates the time step that would be given by the hydro CFL. All calculations were performed on a grid made of 1024 cells. 
In the STS calculations, for each value of $\Delta t$, calculations have been performed  using various values of $N_\mathrm{sts}$ and $\nu_\mathrm{sts}$. For the CG method, only the convergence criterion 
$\epsilon_\mathrm{conv}$ changes.

Figure \ref{sts_cg_fig} shows the radiative energy profiles at time $t=1\times10^{-13}$ s for all calculations we  performed. In all panels, the analytic solution is plotted (black line). The two upper plots give results for the CG method. For $\Delta t\geq 10^{-14}$, the accuracy is very limited. We also see that for $\Delta t\geq 10^{-13}$, the diffusion wave does not propagate at the right speed. The total energy is conserved, but the diffusion wave does not have the correct extent. 
But the STS results are much more accurate, except for $N_\mathrm{sts}=20$ and $\nu_\mathrm{sts}=1\times 10^{-6}$. By construction, STS is expected to be more accurate. The stability is poor when $N_\mathrm{sts}=20$ and $\nu_\mathrm{sts}=1\times 10^{-6}$ because $\nu_\mathrm{sts}$ is close to the stability limit \citep[see][]{Alexiades_1996}. 
\begin{figure}[htb]
  \centering
  \includegraphics[width=9cm,height=7cm]{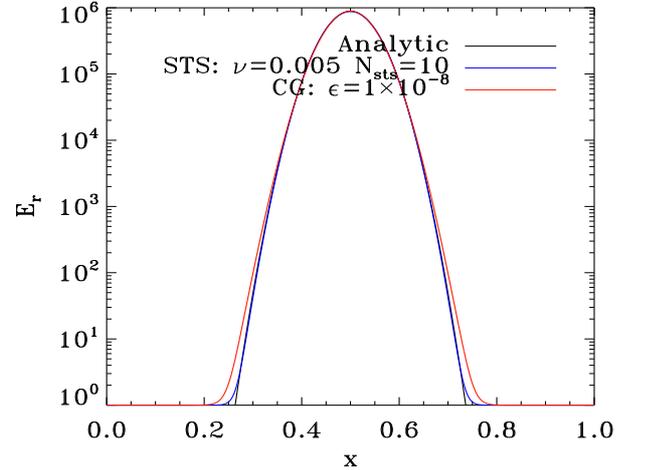}
  \caption{Comparison  of calculations done using  STS or CG and a variable time step given by $\Delta t=1\times10^{-16}*1.05^\mathrm{istep}$, where istep is the index of the number of global (hydro) time steps. Results are given at time $t=1\times 10^{-13}$s.}
  \label{var_dt}
\end{figure}

\begin{figure}[htb]
  \centering
  \includegraphics[width=9cm,height=9cm]{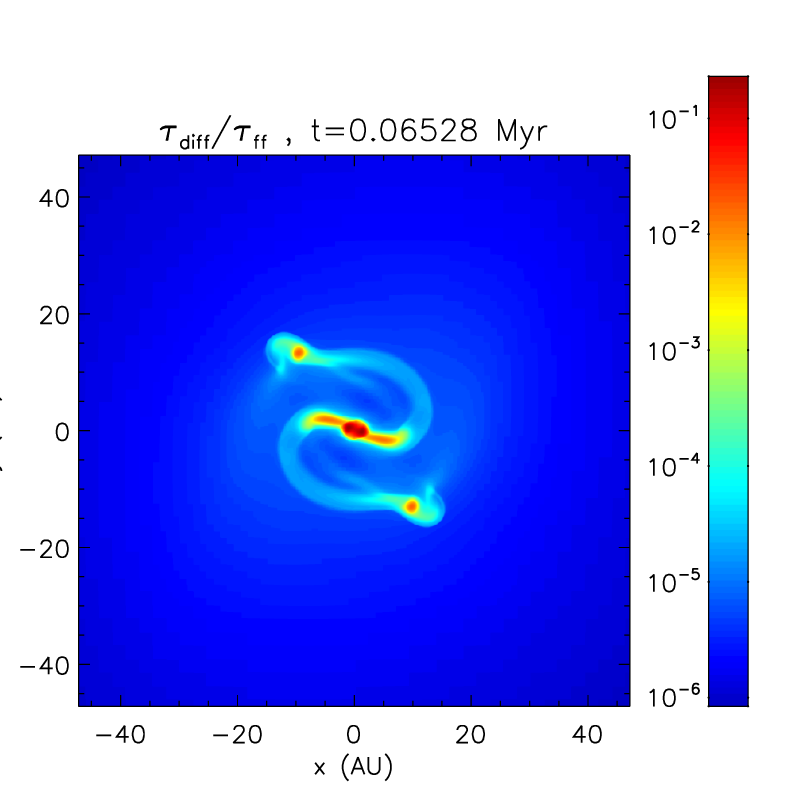}
  \caption{Contours in the equatorial plane of the ratio between diffusion and free-fall times for collapse calculations. The diffusion time is estimated as $\tau_\mathrm{diff}=\frac{l^2}{\frac{c}{3\kappa_\mathrm{R}\rho}}$, where $l$ is the local Jeans length.}
  \label{tdiff_tff}
\end{figure}
\begin{table*}[htb]
\center
\begin{tabular}{|c|c|c|c|c|}
\hline
 Method & Parameters & $\Delta t$ & CPU time (s) & $N_\mathrm{iter}$ \\
\hline
\hline
  CG &  $\epsilon_\mathrm{conv}=1\times10^{-6}$                  & $1\times10^{-16}$  & 82.9 & 10805 \\
  CG &  $\epsilon_\mathrm{conv}=1\times10^{-8}$                  & $1\times10^{-16}$  & 68.5 & 14623 \\
 STS &  $\nu_\mathrm{sts}=0.001 $ , N$_\mathrm{sts}=10$          & $1\times10^{-16}$  & 20.4 &  9000 \\
\hline 
  CG &  $\epsilon_\mathrm{conv}=1\times10^{-6}$                  & $1\times10^{-15}$  & 21.2 &  4738 \\
  CG &  $\epsilon_\mathrm{conv}=1\times10^{-8}$                  & $1\times10^{-15}$  & 27.8 &  6456 \\
 STS &  $\nu_\mathrm{sts}=1\times10^{-6} $ , N$_\mathrm{sts}=20$ & $1\times10^{-15}$  &  2.8 &  2107 \\
 STS &  $\nu_\mathrm{sts}=0.001 $ , N$_\mathrm{sts}= 5$          & $1\times10^{-15}$  &  2.9 &  2408 \\
 STS &  $\nu_\mathrm{sts}=0.001 $ , N$_\mathrm{sts}=20$          & $1\times10^{-15}$  &  2.9 &  2408 \\
 STS &  $\nu_\mathrm{sts}=0.001 $ , N$_\mathrm{sts}=10$          & $1\times10^{-15}$  &  2.9 &  2408 \\
 STS &  $\nu_\mathrm{sts}=0.005 $ , N$_\mathrm{sts}= 5$          & $1\times10^{-15}$  &  3.1 &  2709 \\
 STS &  $\nu_\mathrm{sts}=0.005 $ , N$_\mathrm{sts}=10$          & $1\times10^{-15}$  & 3.04 &  2709 \\
\hline
  CG &  $\epsilon_\mathrm{conv}=1\times10^{-6}$                  & $1\times10^{-14}$  & 11.4 &  2848 \\
  CG &  $\epsilon_\mathrm{conv}=1\times10^{-8}$                  & $1\times10^{-14}$  & 15.4 &  3892 \\
 STS &  $\nu_\mathrm{sts}=1\times10^{-6} $ , N$_\mathrm{sts}=20$ & $1\times10^{-14}$  &  0.6 &   600 \\
 STS &  $\nu_\mathrm{sts}=0.001 $ , N$_\mathrm{sts}= 5$          & $1\times10^{-14}$  & 0.99 &  1470 \\
 STS &  $\nu_\mathrm{sts}=0.001 $ , N$_\mathrm{sts}=20$          & $1\times10^{-14}$  & 0.75 &   900 \\
 STS &  $\nu_\mathrm{sts}=0.001 $ , N$_\mathrm{sts}=10$          & $1\times10^{-14}$  & 0.69 &   780 \\
 STS &  $\nu_\mathrm{sts}=0.005 $ , N$_\mathrm{sts}= 5$          & $1\times10^{-14}$  &  1.1 &  1620 \\
 STS &  $\nu_\mathrm{sts}=0.005 $ , N$_\mathrm{sts}=10$          & $1\times10^{-14}$  & 0.87 &  1170 \\
\hline
  CG &  $\epsilon_\mathrm{conv}=1\times10^{-6}$                  & $5\times10^{-14}$  &  6.9 &  1755 \\
  CG &  $\epsilon_\mathrm{conv}=1\times10^{-8}$                  & $5\times10^{-14}$  &  9.3 &  2365 \\
 STS &  $\nu_\mathrm{sts}=1\times10^{-6} $ , N$_\mathrm{sts}=20$ & $5\times10^{-14}$  & 0.39 &   390 \\
 STS &  $\nu_\mathrm{sts}=0.001 $ , N$_\mathrm{sts}= 5$          & $5\times10^{-14}$  & 0.8  &  1326 \\
 STS &  $\nu_\mathrm{sts}=0.001 $ , N$_\mathrm{sts}=20$          & $5\times10^{-14}$  & 0.56 &   756 \\
 STS &  $\nu_\mathrm{sts}=0.001 $ , N$_\mathrm{sts}=10$          & $5\times10^{-14}$  & 0.46 &   534 \\
 STS &  $\nu_\mathrm{sts}=0.005 $ , N$_\mathrm{sts}= 5$          & $5\times10^{-14}$  & 0.87 &  1476 \\
 STS &  $\nu_\mathrm{sts}=0.005 $ , N$_\mathrm{sts}=10$          & $5\times10^{-14}$  & 0.68 &  1032 \\
\hline
  CG &  $\epsilon_\mathrm{conv}=1\times10^{-6}$                  & $1\times10^{-13}$  &  4.6 &  1135 \\
  CG &  $\epsilon_\mathrm{conv}=1\times10^{-8}$                  & $1\times10^{-13}$  &  5.6 &  1399 \\
 STS &  $\nu_\mathrm{sts}=1\times10^{-6} $ , N$_\mathrm{sts}=20$ & $1\times10^{-13}$  & 0.36 &   351 \\
 STS &  $\nu_\mathrm{sts}=0.001 $ , N$_\mathrm{sts}= 5$          & $1\times10^{-13}$  & 0.77 &  1311 \\
 STS &  $\nu_\mathrm{sts}=0.001 $ , N$_\mathrm{sts}=20$          & $1\times10^{-13}$  & 0.52 &   729 \\
 STS &  $\nu_\mathrm{sts}=0.001 $ , N$_\mathrm{sts}=10$          & $1\times10^{-13}$  & 0.43 &   495 \\
 STS &  $\nu_\mathrm{sts}=0.005 $ , N$_\mathrm{sts}= 5$          & $1\times10^{-13}$  & 0.83 &  1467 \\
 STS &  $\nu_\mathrm{sts}=0.005 $ , N$_\mathrm{sts}=10$          & $1\times10^{-13}$  & 0.65 &  1020 \\
\hline
\hline
  CG &  $\epsilon_\mathrm{conv}=1\times10^{-8}$                  & $1\times10^{-16}*1.05^\mathrm{istep}$  &  19 &  4680 \\
 STS &  $\nu_\mathrm{sts}=0.005 $ , N$_\mathrm{sts}= 10$         & $1\times10^{-16}*1.05^\mathrm{istep}$  &  1.7 &  1782 \\
\hline
\end{tabular} 
\caption{Summary of calculations plotted in Fig. \ref{sts_cg_fig}. CPU time, the number of iterations in the CG method or the number of substeps in the STS method are given for various time steps and various values of $\epsilon_\mathrm{conv}$ for the CG, and N$_\mathrm{sts}$ and $\nu_\mathrm{sts}$ for STS. }
\label{summary_sts}
\end{table*}

In Table \ref{summary_sts} we give the CPU time and $N_\mathrm{iter}$, which corresponds to the number of iterations for the CG and to the number of substeps for the STS. The number of operations per iteration in the CG and per substep in the STS are equivalent, since it involves the same number of cells (1024). The CPU time spent with the STS is ten times  smaller than the one of the CG method. The STS also requires often twice less iterations than the CG.  The bottom lines give the results for calculations made with a variable time step, which increases with time. The corresponding profiles are plotted in Fig. \ref{var_dt}. The STS remains more accurate in this case than the CG, which is quite accurate over more than three orders of magnitude. The CG gives good results, because, thanks to the variable time steps, the diffusion wave propagates at a correct speed. Indeed, at $t=0$, the gradient of radiative energy is steep and the diffusion wave speed is very high. Using an initial short time step $\Delta t=10^{-16}$ enables us to be closer to the CFL condition associated to the diffusion wave speed. Then, radiative energy gradients and the former CFL condition relax and the time step can increase with time. This relaxation on the integration time step enables us to maximise the accuracy of implicit methods using a subcycling scheme based on the diffusion wave speed propagation. However, this speed remains quite difficult to estimate. 

Eventually, we must conclude by pointing out that even if the STS method is well adapted for this problem, it remains very limited for star-formation calculations. Indeed, the diffusion time is very short compared to the dynamical time estimated as the free-fall time (see Fig. \ref{tdiff_tff}) and then, the STS requires too many substeps. The convergence of the CG depends on the nature of the problem and is not affected by strong differences between the diffusion and the dynamical times. Moreover, we never encounter these steep gradients in the radiative energy  distribution in star-formation calculations. The STS could be efficient only within the fragments, where the diffusion time is very long. This is the reason why we only use  the CG method throughout. An alternative but non-trivial solution would be to couple the CG and the STS methods.

\end{appendix}
\end{document}